\documentclass[prl,twocolumn,showpacs,floatfix,superscriptaddress,a4paper]{revtex4}

\usepackage[tbtags]{amsmath}
\usepackage{graphicx,amssymb,bm}

\newcommand{\be}{\begin{equation}}
\newcommand{\ee}{\end{equation}}
\newcommand{\Tr}{\mathop\mathrm{Tr}\nolimits}

\newcommand{\Dutchij}{i\kern -0.08em j}

\newcommand{\To}{\rightarrow}
\DeclareMathOperator{\re}{Re}
\DeclareMathOperator{\im}{Im}

\begin{document}

\title{A quantum effect in the classical limit:\\ nonequilibrium tunneling in the Duffing oscillator}

\author{Shu-Hao Yeh}
\affiliation{%
Research Center for Applied Sciences, Academia Sinica, 128 Sec.~2 Academia Rd., Nankang, Taipei 11529, Taiwan}
\affiliation{%
Department of Chemistry, Purdue University, West Lafayette, IN 47907, USA}

\author{Dong-Bang Tsai}
\affiliation{%
Research Center for Applied Sciences, Academia Sinica, 128 Sec.~2 Academia Rd., Nankang, Taipei 11529, Taiwan}
\affiliation{%
Department of Applied Physics, Stanford University, Stanford, CA 94305, USA}

\author{Che-Wei Huang}
\affiliation{%
Research Center for Applied Sciences, Academia Sinica, 128 Sec.~2 Academia Rd., Nankang, Taipei 11529, Taiwan}

\author{Md.\ Manirul Ali}
\affiliation{%
Research Center for Applied Sciences, Academia Sinica, 128 Sec.~2 Academia Rd., Nankang, Taipei 11529, Taiwan}

\author{Alec \surname{Maassen van den Brink}}
\email{alec@gate.sinica.edu.tw}
\affiliation{%
Research Center for Applied Sciences, Academia Sinica, 128 Sec.~2 Academia Rd., Nankang, Taipei 11529, Taiwan}
\affiliation{%
Department of Physics, National Cheng Kung University, Tainan 70101, Taiwan}

\date{\today}

\begin{abstract}
For suitable parameters, the classical Duffing oscillator has a well-understood bistability in its stationary states, with low- and high-amplitude branches connected by an unstable intermediate branch. As expected from the analogy with a particle in a double-well potential, transitions between these states become possible either at finite temperature due to noise, or in the quantum regime due to tunneling. In this analogy, besides local stability to small perturbations, one can also discuss global stability by comparing the two potential minima. For the Duffing oscillator, the stationary states emerge dynamically from an interplay between driving and dissipation so that \textit{a priori}, a potential-minimum criterion for them does not exist locally, let alone globally. However, global stability is still relevant, and definable as the state containing the majority (in the limit, all) population for long times, low temperature, and close to the classical limit. Further, the \emph{crossover point} is the parameter value at which global stability abruptly changes from one state to the other.

For the double-well model, the crossover point is unambiguously defined by potential-minimum degeneracy. Given that this analogy is so effective in other respects, it is thus striking that for the Duffing oscillator, the crossover point turns out to be non-unique. Rather, none of the three aforementioned limits commute with each other, and the limiting behaviour depends on the order in which they are taken. More generally, as both $\hbar\To0$ and $T\To0$, the ratio $\hbar\omega_0/k_\mathrm{B}T$ continues to be a key parameter and can have any nonnegative value. This points to an apparent conceptual difference between equilibrium and nonequilibrium tunneling. We present numerical evidence for this phenomenon by studying the pertinent quantum master equation in the photon-number basis. Independent verification and some further understanding are obtained using a semi-analytical approach in the coherent-state representation.
\end{abstract}

\pacs{03.65.Yz
, 05.40.-a
, 85.25.Cp
}

\maketitle

\paragraph{Introduction.}

The Duffing model describes an oscillator with weak damping, driving, and a quartic potential nonlinearity~\cite{pedia}. Given that a cubic nonlinearity renders the potential unbounded from below on one side, with essentially different physics, and that the quartic term is the next one encountered in the expansion of a generic one-dimensional potential, the model is a prototype for weakly anharmonic oscillators. As such, it is unsurprising that it is encountered when describing a wide range of oscillation phenomena, be they mechanical, electric, or optical. For a range of parameters (discussed in detail below), the model exhibits a well-known classical bistability. This leads to the application serving as our main motivation, as a superconducting readout device for superconducting quantum bits (qubits)~\cite{nori}. In this context, the Duffing oscillator is known as the Josephson bifurcation amplifier (JBA)~\cite{JBA}. A key advantage of the JBA is that both bistable stationary states are superconducting, so that one avoids the power dissipation and quasiparticle generation which would accompany a switch to the normally conducting state. However, our focus will be on the oscillator proper, not on its coupling to a qubit or other device.

In the JBA application, the Duffing oscillator is sufficiently large to be in the classical regime, and a classical description suffices for the main experimental results. Nonetheless, a fundamentally consistent treatment of the coupling to a quantum device necessitates a quantum-mechanical description of the entire system~\cite{saito}. Further, states which are (bi)stable classically may acquire a finite lifetime due to quantum \emph{tunneling}, which should be properly quantified even if it is negligible in a given experiment~\cite{dykman,XQL,peano}. More ambitiously, one could try to design an experiment focusing on this unconventional form of \emph{nonequilibrium tunneling} in a driven system, which will indeed turn out to have some unusual and novel features.

First, we will review the classical situation. Subsequently, a quantum formulation is developed and analyzed numerically, introducing and characterizing the phenomenon of \emph{crossover}. As in other contexts, quantum tunneling in the Duffing oscillator can be contrasted with thermal activation and diffusion at finite temperature, which persists in, though is not restricted to, the classical limit. Comparison of the two approaches reveals a striking difference in crossover behaviour in the classical, low-temperature limit. Therefore, we focus on this limit using semi-analytical asymptotic methods. In particular, it is physically transparent (though not convenient numerically) to cast the quantum equations in a form closely resembling the classical ones, using a coherent-state representation. We finish by making some concluding remarks.

\paragraph{Classical dynamics.}

For background, and to establish some notation, let us recapitulate the classical situation. The Newton--Duffing equation is
\be
  \ddot{x}+\omega_0^2x=-\epsilon[\gamma\dot{x}+\alpha x^3+f\cos(\omega t)]\;,\label{ND}
\ee
in which the driving strength~$f$, the damping~$\gamma$, and the potential nonlinearity~$\alpha$ [all defined such that the mass~$m$ cancels in Eq.~(\ref{ND})] are governed by the same small parameter~$\epsilon\To0$. Due to the oscillator's high quality, nonperturbative physics can nonetheless emerge for long times if the driving is sufficiently close to resonance,
\be
  \omega^2-\omega_0^2=\epsilon\Omega\;,
\ee
with $\Omega$ denoting the detuning. Since the potential anharmonicity is $\frac{1}{4}\epsilon m\alpha x^4$, we take $\alpha>0$ to have a globally stable potential~\cite{alphanote}.

\begin{figure}
  \includegraphics[width=9.5cm]{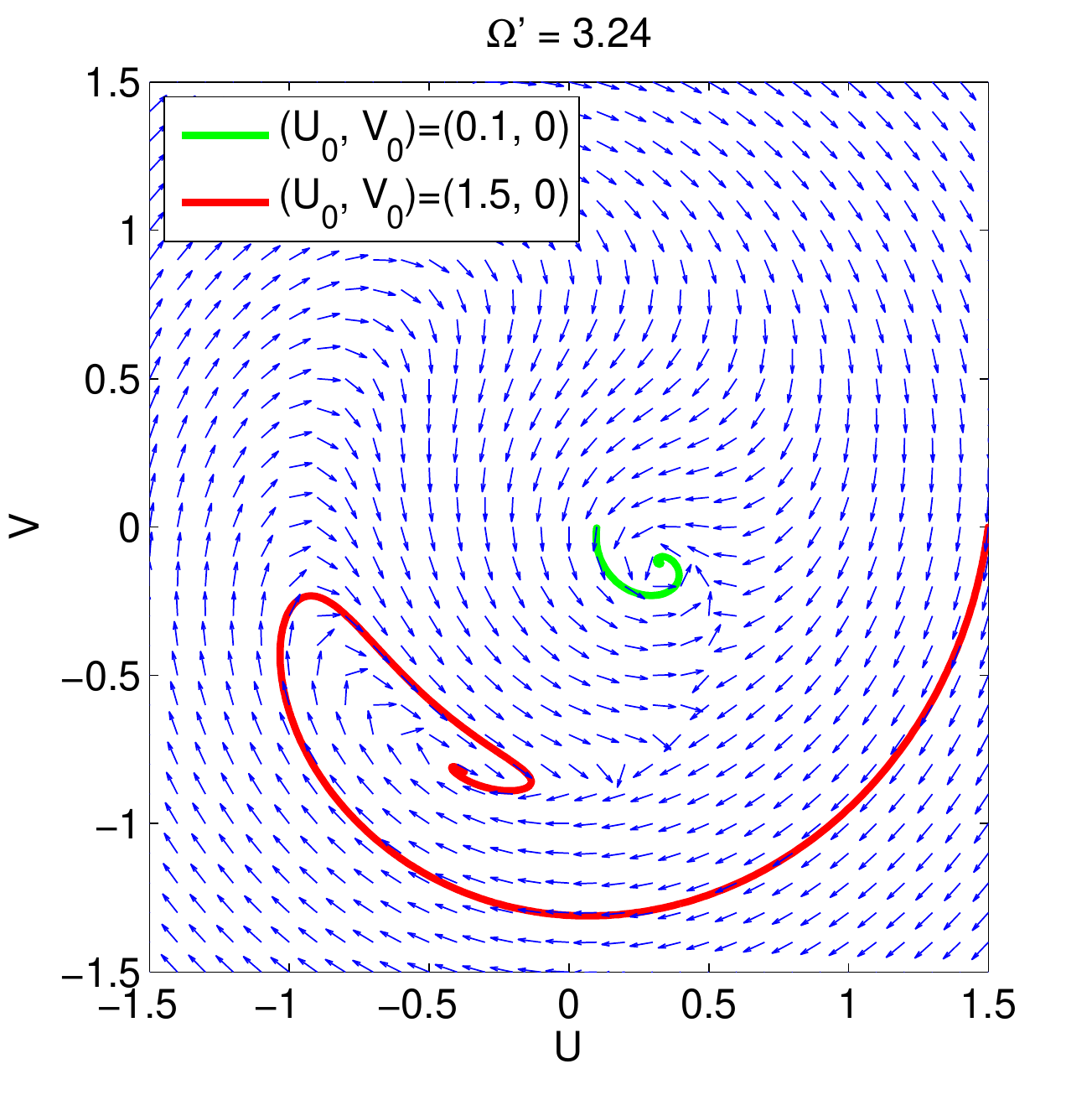}
  \caption{The direction of the $\bm{J}$-flow given by Eq.~(\ref{Jflow}) in the bistable regime. Also shown are two sample integral curves, one (in green) with initial conditions causing it to spiral into the low-amplitude attractor, and one (red) which ends up in the high-amplitude stationary state. The third irregularity in the flow, which can be discerned near $(0.500,-0.496)$, corresponds to the saddle point. Integrating the flow backwards in two directions starting out from the saddle, one obtains the separatrix demarcating the two basins of attraction, which spirals out to infinity in an appealing Yin--Yang pattern, shown in Fig.~\ref{FP}. Physically, while all initial states near the origin are attracted to the low-amplitude state, the fate of initial states with large initial amplitude depends on their phase, so that the two basins of attraction continue to be interleaved for arbitrarily large radii.}
  \label{Jfig}
\end{figure}

To focus on the long-time dynamics, transform the fast ($\sim\omega_0^{-1}$), large, near-harmonic oscillations away by going to the rotating frame, in this context also known as Van der Pol coordinates~\cite{VdP}:
\be
  \begin{pmatrix} u \\ v \end{pmatrix} =
   \begin{pmatrix} \cos\omega t & -\sin\omega t \\
                   \sin\omega t & \hphantom{-}\cos\omega t \end{pmatrix}
   \begin{pmatrix} x \\ p/\omega m \end{pmatrix}\;,\label{uv}
\ee
with $p=m\dot{x}$. One finds that the expressions for $(\dot{u},\dot{v})$ indeed are $\mathcal{O}(\epsilon)$, so that $(u,v)$ are varying slowly; thus, the integrated effect of terms such as $u\cos(\omega t)$ in the equations of motion is of higher order in~$\epsilon$, hence negligible. After this coarse-graining, the surviving slow dynamics is expressed in its natural (dissipative) time scale $\tau\equiv\epsilon\gamma t$, as
\be
  \frac{d}{d\tau}\begin{pmatrix} U \\ V \end{pmatrix} =
  \begin{pmatrix} J_U \\ J_V \end{pmatrix}\;,\label{coarse}
\ee
with the scaled rotating-frame flow field (see Fig.~\ref{Jfig})
\be
  \bm{J}=\frac{1}{2}\begin{pmatrix}
  -\Omega'V-U+\frac{3}{4}\alpha'V(U^2+V^2)\\[2mm]
  \Omega'U-V-1-\frac{3}{4}\alpha'U(U^2+V^2)\end{pmatrix}\;,
  \label{Jflow}
\ee
and the dimensionless variables~\cite{omega0}
\be
  \Omega'=\frac{\Omega}{\omega_0\gamma}\;,\quad
  \alpha'=\frac{\alpha f^2}{\omega_0^3\gamma^3}\;,\quad
  \begin{pmatrix} U \\ V \end{pmatrix} =
  \frac{\omega_0\gamma}{f}\begin{pmatrix} u \\ v \end{pmatrix}\;.
  \label{dim-less}
\ee
In particular, detuning is scaled relative to the natural linewidth, while coordinates are relative to the resonant amplitude.

Unlike the driven Eq.~(\ref{ND}), which is equivalent to an autonomous three-dimensional system and is known to have a chaotic regime~\cite{duffing-chaos}, Eq.~(\ref{coarse}) is an autonomous two-dimensional dynamical system, and according to a theorem by Poincar\'e~\cite{limitcycle} it cannot exhibit such behaviour. The difference is of course the specialization to infinitesimal~$\epsilon$ in the latter.

\begin{figure}
  \includegraphics[width=9.5cm]{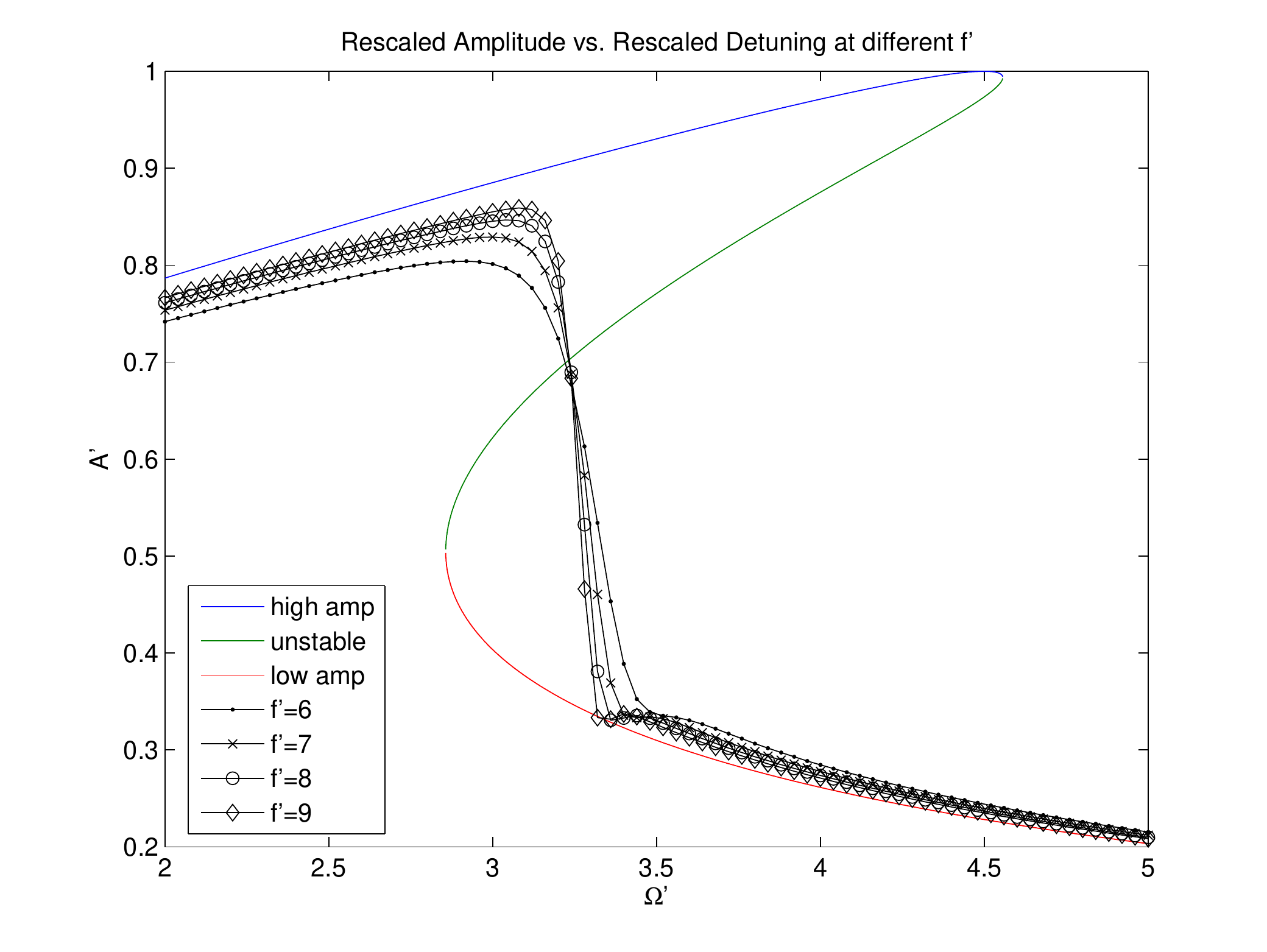}
  \caption{The reduced amplitude response of the Duffing oscillator. The low- (red) and high-amplitude (blue) stationary states, as well as the unstable branch (green), are the classical response~(\ref{classical-A}). Data points denote the stationary quantum response at $T=0$, see below Eq.~(\ref{def-Z}); the stronger the system is driven, the closer it is to the classical limit, and the steeper the curves are near the crossover point $\Omega'\approx3.3$.}
  \label{A-response}
\end{figure}

The stationary states, which were limit cycles~\cite{limitcycle} of the original Eq.~(\ref{ND}), become fixed points in the rotating frame upon coarse-graining, and can be found by setting the right-hand side (rhs) of Eq.~(\ref{coarse}) to zero. In polar coordinates $u=A\cos\phi$, $v=A\sin\phi$, one obtains~\cite{phi-sign}
\begin{eqnarray}
  1&=&{A'}^2+\Bigl(\frac{3}{4}\alpha'{A'}^2-\Omega'\Bigr)^{\!2}{A'}^2\;,\label{classical-A}\\
  \cot\phi&=&\frac{3}{4}\alpha'{A'}^2-\Omega'\;,\label{classical-phi}
\end{eqnarray}
where $A'=A\omega_0\gamma/f$ is defined similarly to $U,V$ in Eq.~(\ref{dim-less}). For $\alpha'\downarrow0$, these reduce to the Lorentzian response of the damped harmonic oscillator, but even in the nonlinear case, the driving $f$ merely sets a scale for the amplitude~$A$. As a result, the amplitude response $A'(\Omega')$ is a one-parameter family of curves, dependent only on~$\alpha'$. For $\alpha'>0$ ($\alpha'<0$) these curves become asymmetric and skewed to the right (left), until at the critical value $\alpha'=32/3^{5/2}\approx2.0528$ the solutions bifurcate, and become multiple-valued for an increasing range of detunings~$\Omega'$---the bistable window (tri-colored line in Fig.~\ref{A-response}). On the one hand, a choice for $\alpha'$ should exceed the critical value sufficiently to afford a substantial bistable window. On the other, if $\alpha'$ is taken too large, the separation between the two stationary states becomes so large that the quantum tunneling phenomenon discussed below is too weak to be observed. For ease of comparison, in all our examples we take $\alpha'=6$, with bistability for $2.8568\lesssim\Omega'\lesssim4.5563$. Finally, note that the maximum amplitude is always $A'=1$, attained for $\Omega'=\frac{3}{4}\alpha'$, as evidenced in Fig.~\ref{A-response}.

\paragraph{Quantum theory.}

Thus far, the Duffing bistability is analogous to the one of a particle in a double-well potential, say
\be
  U(x)=x^4-bx^2+cx\;.\label{dblwellmodel}
\ee
For $c=0$, the unique minimum $U(x{=}0)=0$ for negative $b$ bifurcates at the critical value $b=0$. For $b>0$, there is an increasing (and in this case, symmetric) range of biases~$c$ for which $U(x)$ features two minima, until the left (right) minimum loses local stability at the lower (upper) end of the bistable $c$-window. Thus, the parameters $b$ and $c$ play the roles of $\alpha'$ and $\Omega'$, respectively. Quantum mechanics in a potential such as (\ref{dblwellmodel}) has several outspoken features. For instance, for $b>0$ the ground state rapidly changes character as a function of $c$ in a narrow range (of the order of the tunneling amplitude, hence exponentially small near the classical limit) around $c=0$, from being localized in the left well, via being a symmetric superposition, to being localized in the right well. We call this phenomenon \emph{crossover}; for $U(x)$ in (\ref{dblwellmodel}), the crossover point $c=0$ is obvious by symmetry, but even for more complicated~$U$, one merely has to find the bias point for which the well minima have the same potential.

Guided by the above analogy, in the quantum case one expects tunneling transitions between the two stationary states to become possible also for the Duffing model. Key questions are to which extent such transitions resemble the (equilibrium) tunneling in a potential such as (\ref{dblwellmodel}) on the one hand, and the classical dynamics of Eqs.~(\ref{ND})--(\ref{classical-phi}) on the other, in particular close to the classical limit. Crossover phenomena are of interest as well, especially since the location of the crossover point does not seem to be \emph{a priori} obvious. In Ref.~\onlinecite{saito}, the focus is on the JBA application of a quantum Duffing oscillator coupled to a qubit. In Refs.~\onlinecite{dykman} and \onlinecite{XQL}, quantum Duffing dynamics was studied near the edge of the bistable window (called the critical case in the former and the bifurcation point in the latter). In Ref.~\onlinecite{peano}, the quantum Duffing oscillator is studied using a superoperator Floquet technique.

Quantum tunneling in driven systems has been studied in many contexts~\cite{driven-tunnel}, but here the very location of the stationary states depends on the damping in leading order (since $A\sim\gamma^{-1}A'$, with $A'$ of order one). As a result, the problem does not have a weak-damping limit [indeed, the evolution equation (\ref{QME}) below has damping coefficients equal to unity in its last two terms], and a description in terms of the density matrix $\rho$ must be used from the outset.

We use a quantum Master equation (QME) formulation. The use of a (Markovian) QME for damped quantum oscillators is not always straightforward, or indeed uncontroversial~\cite{zhang}. However, it is justified here since (a)~the damping $\sim\epsilon\gamma$ is very weak compared to the harmonic motion, and (b)~we consider only the long-time, coarse-grained dynamics. Note further that since also the nonlinearity is $\sim\epsilon$, any cross-effect between it and the dissipation is at least $\mathcal{O}(\epsilon^2)$, and beyond our consideration. This justifies the use of the standard harmonic-oscillator damping terms~\cite{harmonic-damping} also for the Duffing oscillator. Because of the restriction to long times, we work exclusively in the rotating frame
\be
  \tilde{\rho}(t)\equiv U^\dagger(t)\rho(t)U(t)\;,\quad U(t)=e^{-i\omega Nt}\;,\quad
  N=a^\dagger a\;,\label{RWA}
\ee
which are analogous to the classical Eq.~(\ref{uv}). Upon performing the coarse-graining or \emph{rotating-wave approximation} (RWA), one obtains without further ado the QME in Lindblad~\cite{lindblad} form
\begin{align}
  i\frac{d\tilde{\rho}}{d\tau}&=[\tilde{H},\tilde{\rho}]+\frac{i}{2}(\bar{n}{+}1)
  (2a\tilde{\rho}a^\dagger-a^\dagger a\tilde{\rho}-\tilde{\rho}a^\dagger a)\notag\\ &\quad+\frac{i}{2}\bar{n}
  (2a^\dagger\tilde{\rho}a-aa^\dagger\tilde{\rho}-\tilde{\rho}aa^\dagger)\label{QME}\\[3mm]
  &\equiv\tilde{\mathcal{L}}\{\tilde{\rho}\}\;,\label{L}
\end{align}
where we introduced the Liouville superoperator $\tilde{\mathcal{L}}$ and the scaled rotating-frame Hamiltonian
\be
  \tilde{H}=-\frac{\Omega'}{2}N+\frac{f'}{2\sqrt{2}}(a+a^\dagger)
                +\frac{3\alpha'}{8f^{\prime2}}(N^2+N)\;.\label{Hrot}
\ee
Here, the last term $\propto\alpha'$ denotes the potential nonlinearity; the form $N^2+N$ arises by rearranging creation and annihilation operators in the expansion of the potential~$x^4$, retaining only co-rotating terms [cf.\ above Eq.~(\ref{coarse})]. The time scale $\tau$ is as in Eq.~(\ref{coarse}). Compared to the classical case, however, there is an extra parameter: the dimensionless driving
\be
  f'=\frac{f}{\gamma}\sqrt{\frac{m}{\hbar\omega_0}}\;.\label{fprime}
\ee
Namely, while the classical energy can be scaled away, in the quantum case the photon number $N$ is discrete, with $N\gg1$ representing a semiclassical situation while $N\lesssim1$ is the deep quantum limit. This is reflected in the definition~(\ref{fprime}), from which one sees that the classical limit $f'\To\infty$ can be achieved either through strong driving~$f$, taking a large mass~$m$, or letting $\hbar\To0$. We have also introduced the Bose occupation number
\be
  \bar{n}=\frac{1}{e^{\hbar\omega_0/k_BT}-1}\;,\label{bose}
\ee
so that for $T\To0$, only the first (decay-type, $\sim a\tilde{\rho}$) dissipative term survives in Eq.~(\ref{QME}), while the excitation term $\sim a^{\dagger}\tilde{\rho}$ vanishes.

\begin{figure}
  \includegraphics[width=8.5cm]{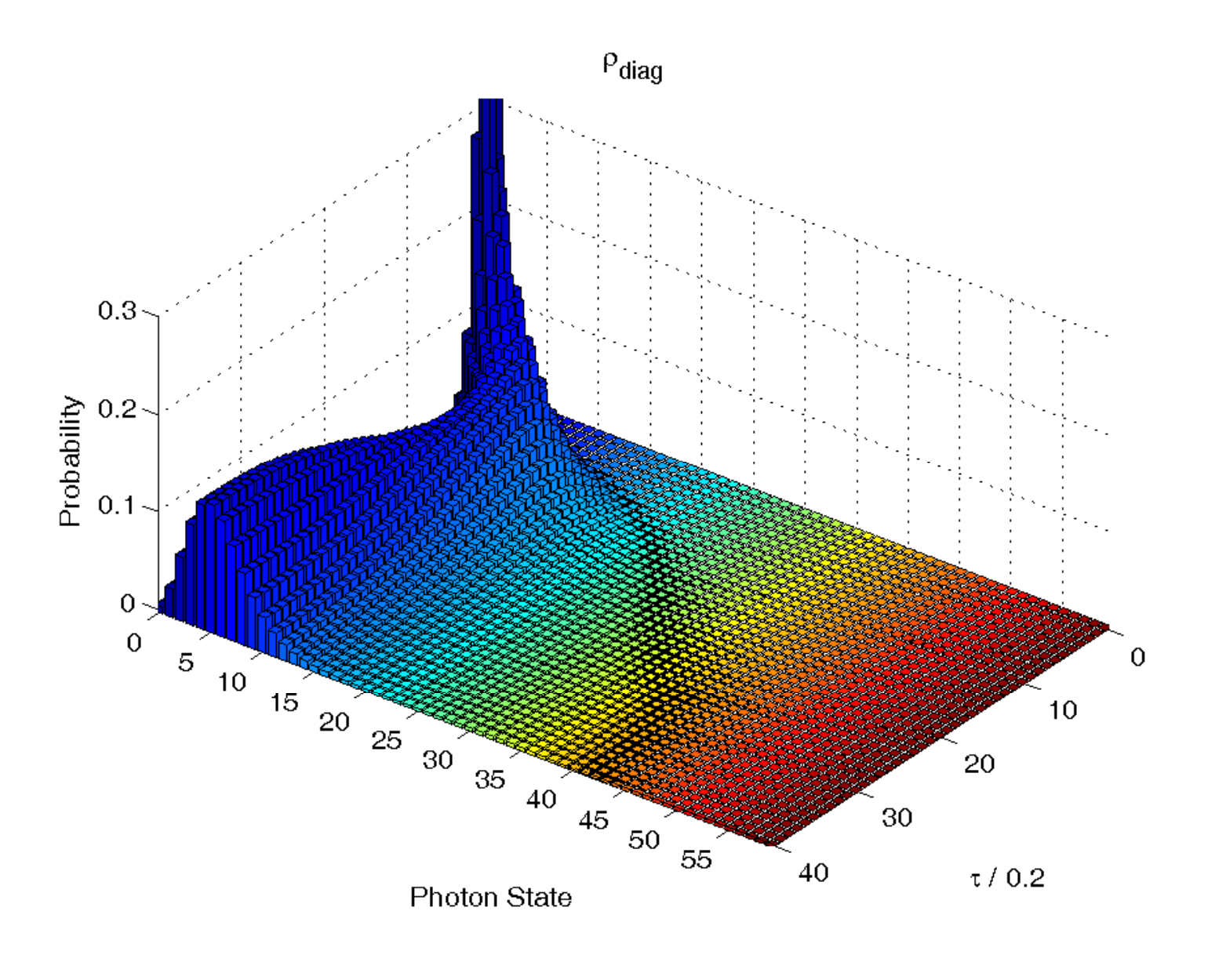}\\
  \includegraphics[width=8.5cm]{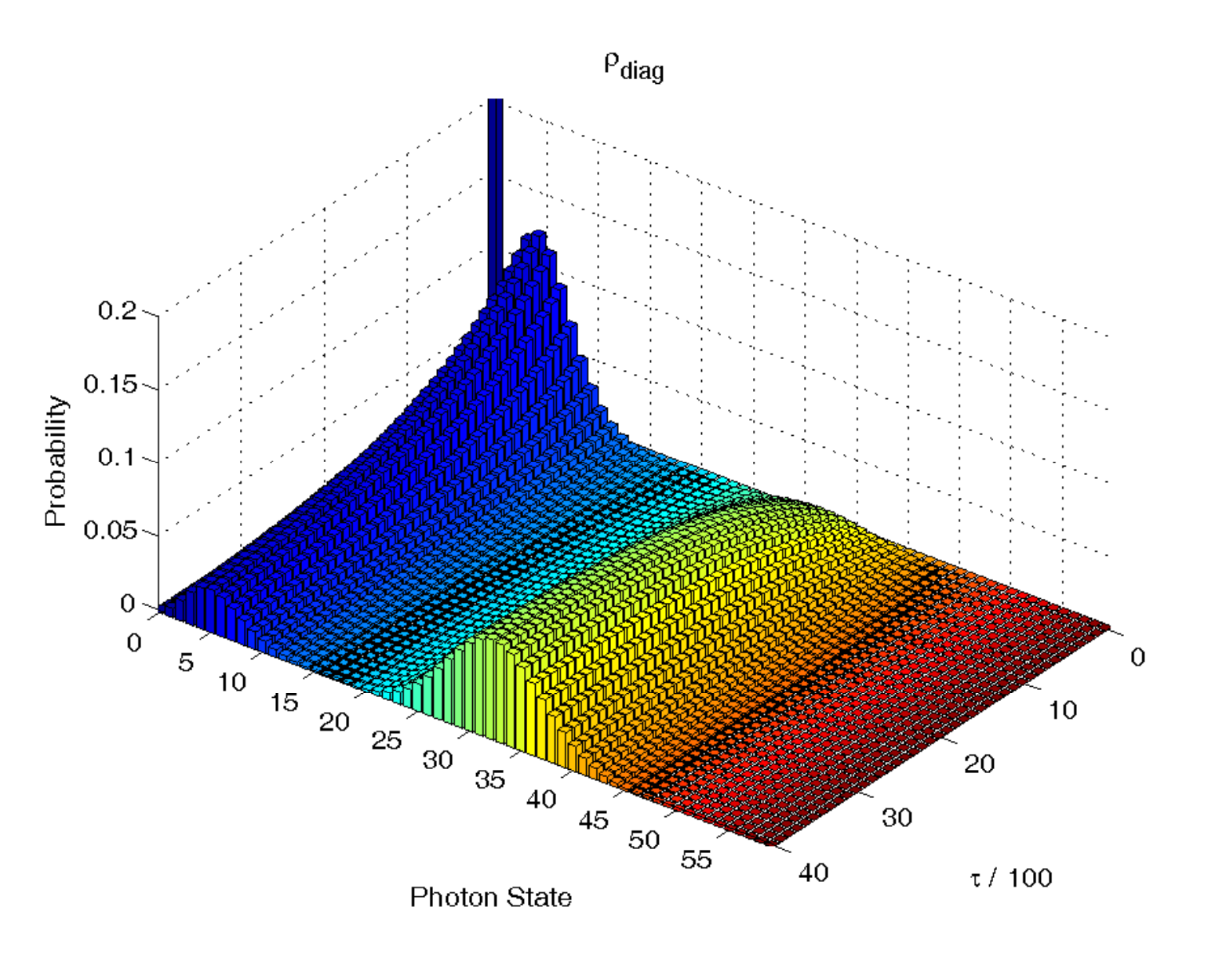}
  \caption{Real-time evolution in the rotating frame on the long (top) and ultralong (bottom) time scale, showing the probability distribution in the number basis as a function of time, for detuning $\Omega'=3.24$ and driving~$f'=7$. While only the diagonal elements of $\langle n|\tilde{\rho}(\tau)|m\rangle$ are thus shown, of course these data have been obtained from a simulation of the full density matrix.}
  \label{real-time}
\end{figure}

For initial exploration, and confirmation of quantum tunneling, Eq.~(\ref{QME}) can be simulated directly by truncating~$N$, choosing a $\tilde{\rho}(\tau{=}0)$, and using a standard RK4 solver for the resulting ODE system. Care has been taken to ensure probability conservation for arbitrary truncation and discretization of the time steps; however, the exponentially small tunneling effects are very sensitive to cutoff artifacts, and convergence must be verified systematically. A representative example is shown in Fig.~\ref{real-time}, starting from the vacuum state $\tilde{\rho}(0)=|0\rangle\langle0|$. Initially, the driving force rapidly increases the photon population, until it becomes counterbalanced by the dissipation (which is more effective for higher states). For times $\tau\sim10$ (top panel), the system seems to settle in a wave-packet type state, which is the quantum counterpart to the classical low-amplitude state (see Fig.~\ref{A-response} for quantitative verification). However, if one persists in simulating until $\tau\sim10^3$ (bottom panel), an entirely new feature emerges in the form of a wave packet corresponding to the classical high-amplitude state, which becomes populated by quantum tunneling from low amplitudes. Only for $\tau\sim10^4$ is a genuine steady state reached (we prefer to avoid the term ``equilibration" for this driven system), through a balance between low-to-high-amplitude tunneling and vice versa. Thus, nonequilibrium tunneling has opened up a new, ultra-long time scale in the quantum Duffing model.

Given that the RWA has yielded a time-independent evolution operator, simulation is ultimately less powerful than directly extracting the $\tau\To\infty$ tunneling dynamics by studying the eigenvalues of $\tilde{\mathcal{L}}$ in Eq.~(\ref{L})~\cite{peano}. We have performed this by numerical diagonalization, using a non-hermitian sparse-matrix routine~\cite{JB}. The main disadvantage of this method is size: while $\tilde{\rho}$ is $N\times N$, the matrix form of $\tilde{\mathcal{L}}$ is $N^2\times N^2$. With some optimization, we have used truncations up to $N\sim120$ and substantially larger $N$ would have been possible, if very time-consuming. While this method thus is slightly laborious, especially close to the classical limit where quantum--classical correspondence must be studied, we have found its excellent stability and reliable convergence to be worth the effort.

As a result, one obtains eigenvalues $\lambda_j$ and eigen\-``vectors" $\tilde{\rho}_j$, in terms of which the time evolution can be expressed as
\be
  \tilde{\rho}(\tau)=\sum_{j\ge1}c_j\tilde{\rho}_je^{-i\lambda_j\tau}\;.
  \label{rho-eigen}
\ee
In contrast with Hamiltonian spectra, only the sum $\tilde{\rho}(\tau)$ is required to be physical, while the individual $\tilde{\rho}_j$ need not have all the properties of density matrices, and can for instance be non-hermitian. From the structure of $\tilde{\mathcal{L}}$ in Eq.~(\ref{QME}), one readily deduces that for each eigenpair $(\lambda_j,\tilde{\rho}_j)$, one also has the pair $(-\lambda_j^*,\tilde{\rho}_j^{\dagger})$, and this symmetry is indeed observed in the numerical output. Thus, the eigenvalues either occur in conjugate pairs, or are purely imaginary. Further, it is essential for stability that no eigenvalues lie in the upper-half complex plane, which would lead to a divergent evolution $\propto e^{\im\lambda_j\tau}$ in Eq.~(\ref{rho-eigen}).

\begin{figure}
  \includegraphics[width=8.5cm]{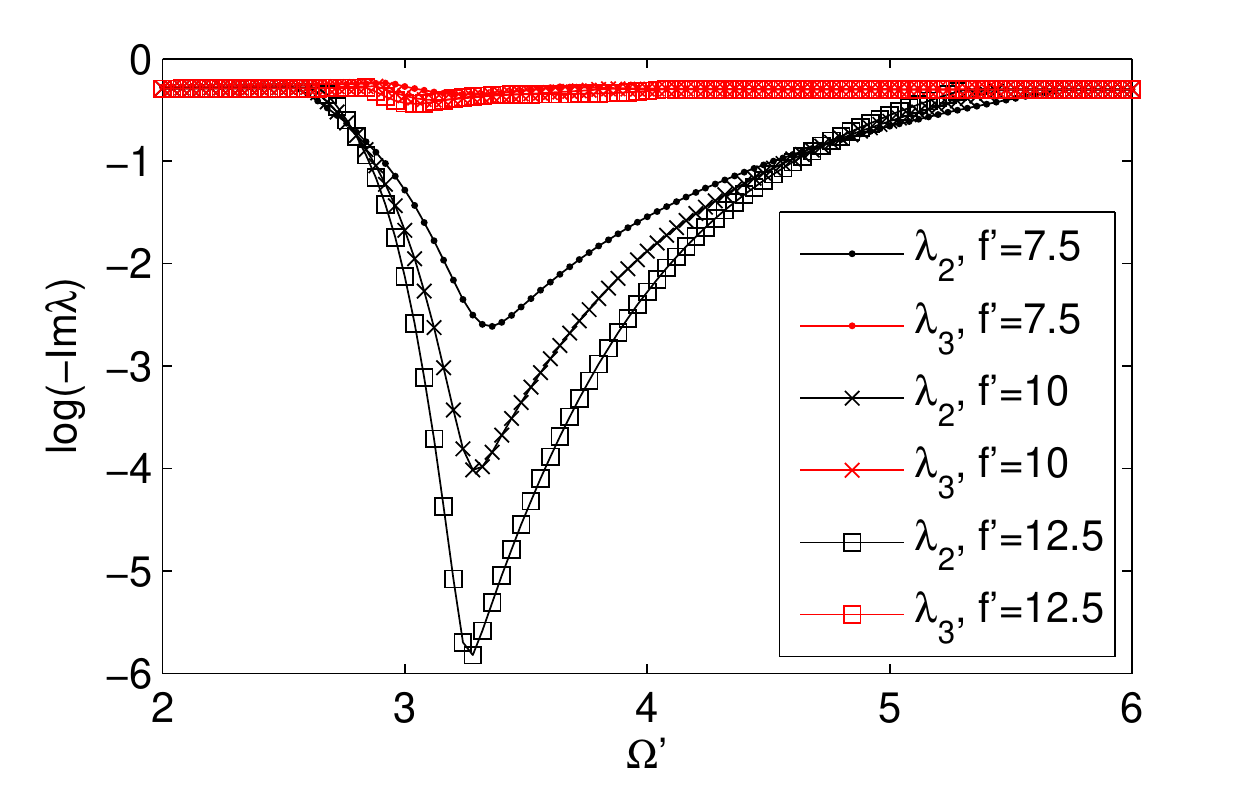}
  \caption{In the bistable region, only the eigenvalue~$\lambda_2$, which is related to tunneling, decays exponentially with the driving strength~$f'$. This behaviour is more pronounced near the crossover point.}
  \label{separation}
\end{figure}

\begin{figure}
  \includegraphics[width=8.5cm]{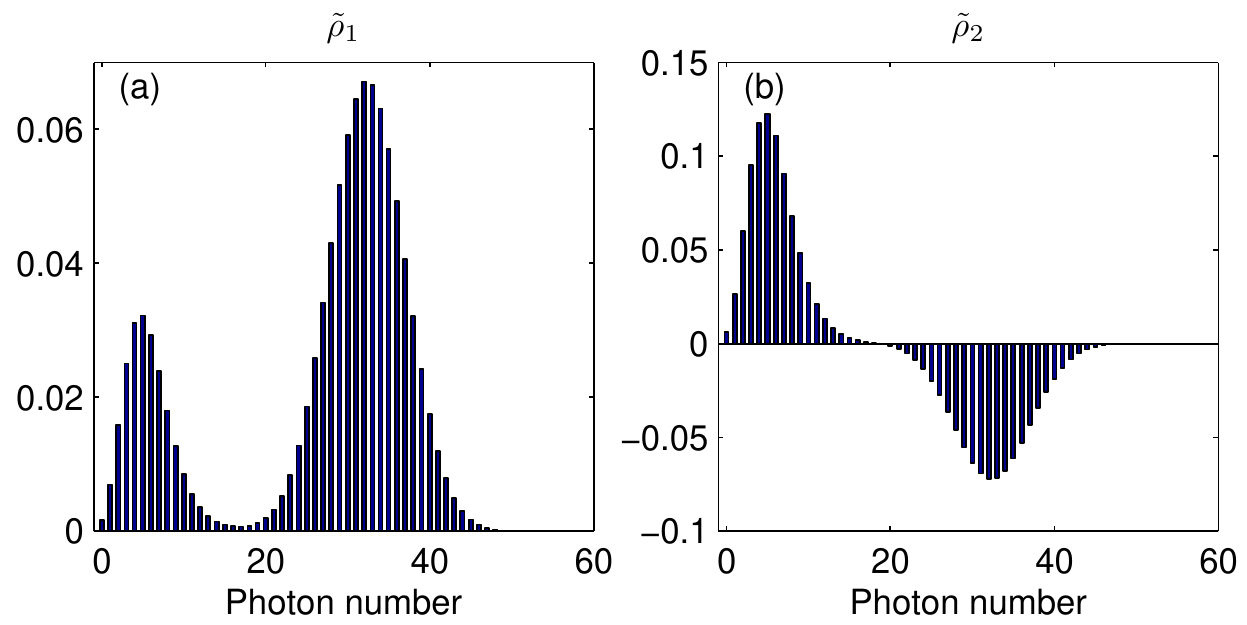}
  \caption{Diagonal elements of the stationary density matrix $\tilde{\rho}_1$~(a) and the tunneling ``density matrix"~$\tilde{\rho}_2$~(b), for detuning $\Omega'=3.24$ and driving~$f'=9$. In the two-state approximation (\ref{twostate}), $\tilde{\rho}_2$ contains no new information, since its peaks are the same shape as those of~$\tilde{\rho}_1$, but are constrained to have equal area and opposite sign (the overall prefactor of $\tilde{\rho}_2$ is arbitrary). Instead, the dynamical information is contained in the eigenvalue~$\lambda_2$, see Fig.~\ref{separation}.}
  \label{rho12}
\end{figure}

We sort the output of the iterative diagonalization by increasing $-\im\lambda_j$; clearly, the more of these one wishes to represent accurately, the more photon-number states must be included. With this convention, we find that in fact $\im\lambda_j<0$ for all~$j$, with the exception of
\be
  \lambda_1=0\;,\label{lambda1}
\ee
representing the unique stationary state. Further, with increasing~$f'$, one observes a separation of time scales illustrated in Fig.~\ref{separation}: while $\lambda_{j\ge3}$ remain of order one, $\lambda_2$ (which thus is purely imaginary) becomes exponentially small, so that it must be carefully resolved from~$\lambda_1$. Physically, on the time scale $\tau\sim1$, the quantum system relaxes towards quantum counterparts of the two classical states within each basin of attraction, analogous to the dynamics shown in Fig.~\ref{Jfig}. On the new, ultra-long time scale $\sim|\lambda_2|^{-1}$, quantum tunneling exchanges population between these two states, until the true stationary state $\tilde{\rho}_1$ is reached for $\tau\To\infty$. As a consequence, for $\tau\gg1$ one has, with exponential accuracy,
\be
  \tilde{\rho}(\tau)\approx\tilde{\rho}_1+c_2\tilde{\rho}_2e^{-|\lambda_2|\tau}\;,
  \label{rho-relax}
\ee
where only $c_2$ depends on the initial conditions. Probability conservation implies
\be
  \Tr\tilde{\rho}_2=0\;,\label{rho2}
\ee
which is indeed satisfied by the numerical result, see Fig.~\ref{rho12}. In practice, Eqs.\ (\ref{lambda1}) and~(\ref{rho2}) serve as a useful test for convergence with respect to the number of included photon states (the details of this convergence are sensitive to the exact manner in which $\tilde{\mathcal{L}}$ is truncated).

\begin{figure}
  \includegraphics[width=8.5cm]{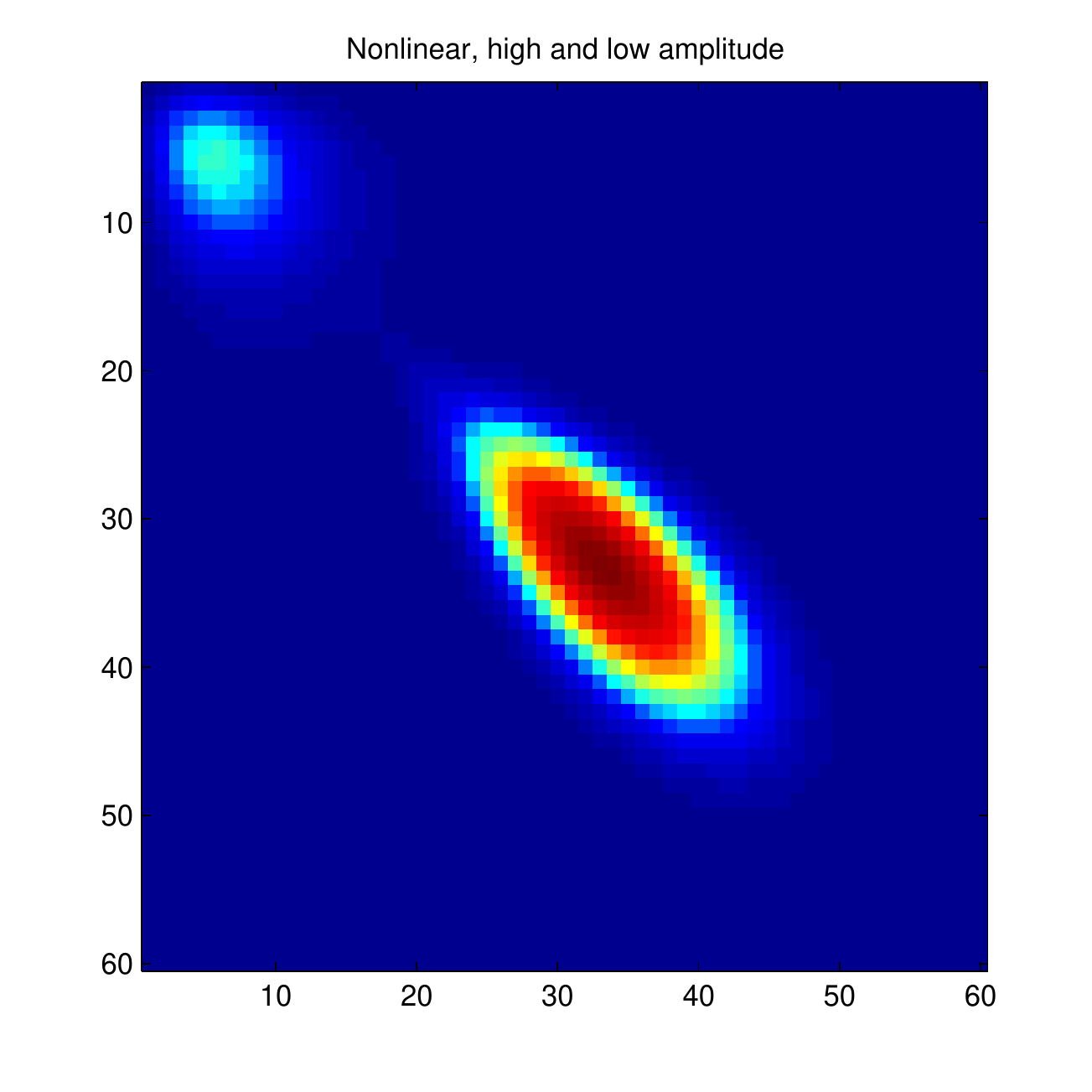}
  \caption{The magnitudes of the stationary density-matrix elements in the number basis $\langle n|\tilde{\rho}_1|m\rangle$, for $\Omega'=3.23$ and~$f'=9$.}
  \label{rho-Nbasis}
\end{figure}

The magnitudes of $\langle n|\tilde{\rho}_1|m\rangle$ are shown in Fig.~\ref{rho-Nbasis}, for parameters in the bistable regime close to the crossover point. The two lobes centered on $(L,L)$ and $(H,H)$ are readily distinguished, with $L\approx6$ and $H\approx33$. Their extent away from the main diagonal shows that $\tilde{\rho}_1$ contains two semi-classical states with significant partial coherence (the quantification of which is possible, but not our priority here). We have examined these same data also on a logarithmic scale, down to machine precision. While the lobes of Fig.~\ref{rho-Nbasis} are then seen to have tails extending far away from the main diagonal, no further structure is observed centered on $(L,H)$ and $(H,L)$. Thus, while each lobe is partially coherent, $\tilde{\rho}_1$ is a mixture of them and not a superposition. Together with $\lambda_2$ being purely imaginary, this supports our claim that \emph{the type of tunneling discussed here is incoherent}---not surprising given the prominent role of dissipation in the Duffing oscillator~\cite{phase-rel}.

\begin{figure}
  \includegraphics[width=8.5cm]{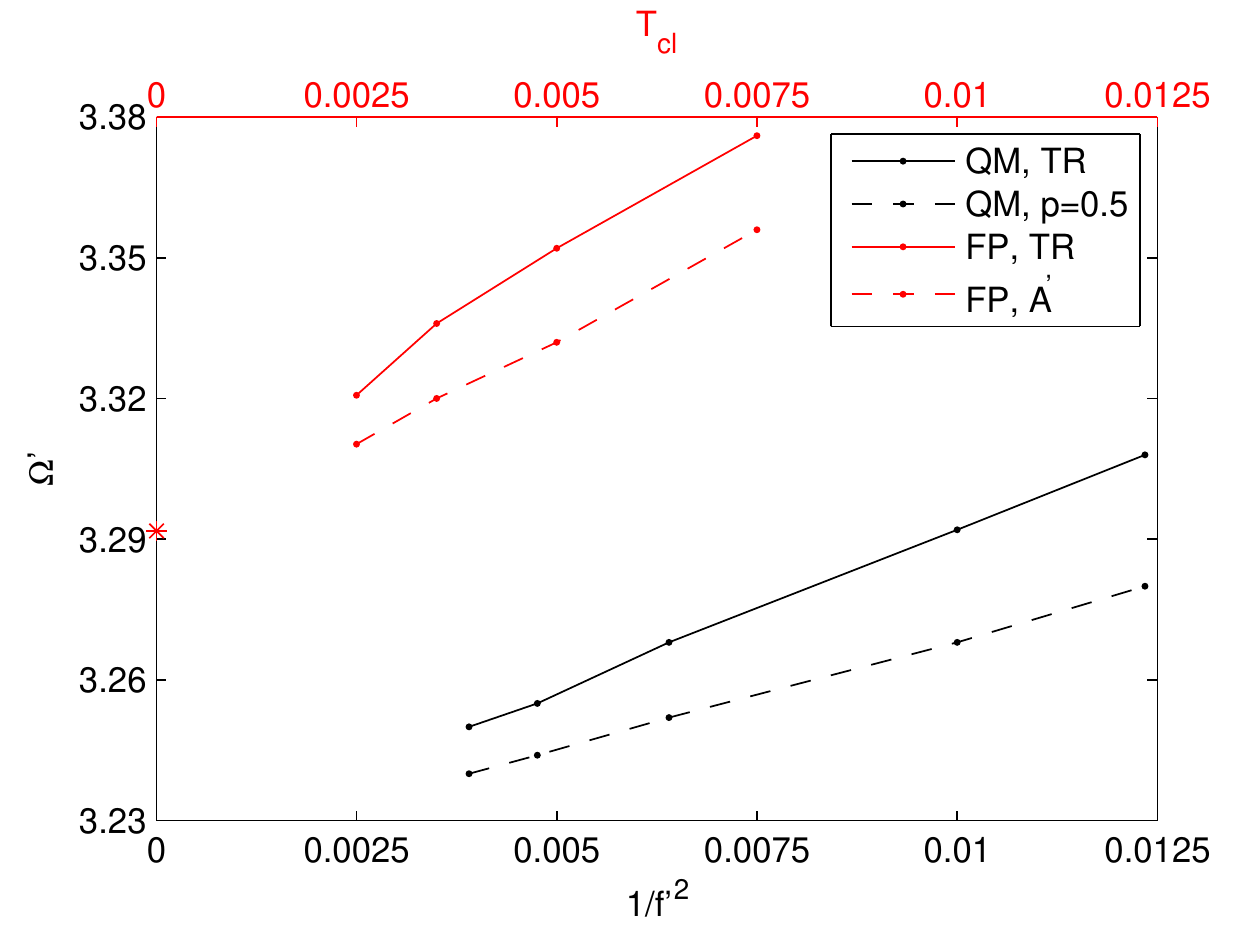}
  \caption{The crossover point of the Duffing model, as extracted from finite-$T$ classical (red) and $T=0$ quantum data (black). In either case, the dashed line links data points obtained using a (static) equiprobability condition, while the solid lines are obtained if one uses the minimum in the (dynamic) relaxation rate $|\lambda_2|$ to pinpoint the crossover. Extrapolating any curve to the $y$-axis realizes one manner of taking the $\hbar,T\To0$ limit. For each color, dashed and solid lines extrapolate to the same $\Omega'$ in the limit of abrupt crossover. However, the red and black curves do not agree with each other. The red star on the $y$-axis refers to the classical crossover point found in Fig.~\ref{crossover-class} using data obtained from Eq.~(\ref{shooting-eqn}), in which the $T\To0$ extrapolation is performed analytically.}
  \label{extrapolate}
\end{figure}

The above implies that, for the longest times, the relaxation (\ref{rho-relax}) is governed by a simple two-state rate equation
\be
  \frac{d}{d\tau}\begin{pmatrix} P_H \\ P_L \end{pmatrix} =
  \begin{pmatrix} -\Gamma_\downarrow & \hphantom{-}\Gamma_\uparrow \\
  \hphantom{-}\Gamma_\downarrow & -\Gamma_\uparrow \end{pmatrix}
  \begin{pmatrix} P_H \\ P_L \end{pmatrix}\;,\label{twostate}
\ee
where $\Gamma_\downarrow$ denotes the transition rate from the high-amplitude semiclassical state to the low-amplitude one, etc. Equation~(\ref{twostate}) implies $\bar{P}_H=1-\bar{P}_L=\Gamma_\uparrow/(\Gamma_\uparrow{+}\Gamma_\downarrow)$ for the stationary distribution and $|\lambda_2|=\Gamma_\uparrow+\Gamma_\downarrow$ for the relaxation rate. It follows that $\lambda_2$ is dominated by the \emph{larger} of the two transition rates, so that it is expected to be small only in the middle of the bistable window. Namely, near the boundary, the upper well in the potential-tunneling analogue (\ref{dblwellmodel}) becomes shallow, so that the corresponding escape rate is large; see Fig.~\ref{separation}. Thus, $|\lambda_2|$ should attain its minimum for, or at least close to, $\Gamma_\uparrow=\Gamma_\downarrow$, in which case $\bar{P}_H=\bar{P}_L=\frac{1}{2}$---the crossover point. Properly, the model (\ref{twostate}) should be regarded as phenomenological, and $\tilde{\rho}_1$ and $|\lambda_2|$ are independent numerical data. Either equiprobability for the former or the minimum in the latter can be used to pinpoint the crossover, with slightly different but consistent results (black lines in Fig.~\ref{extrapolate}).

\begin{figure}
  \includegraphics[width=8.5cm]{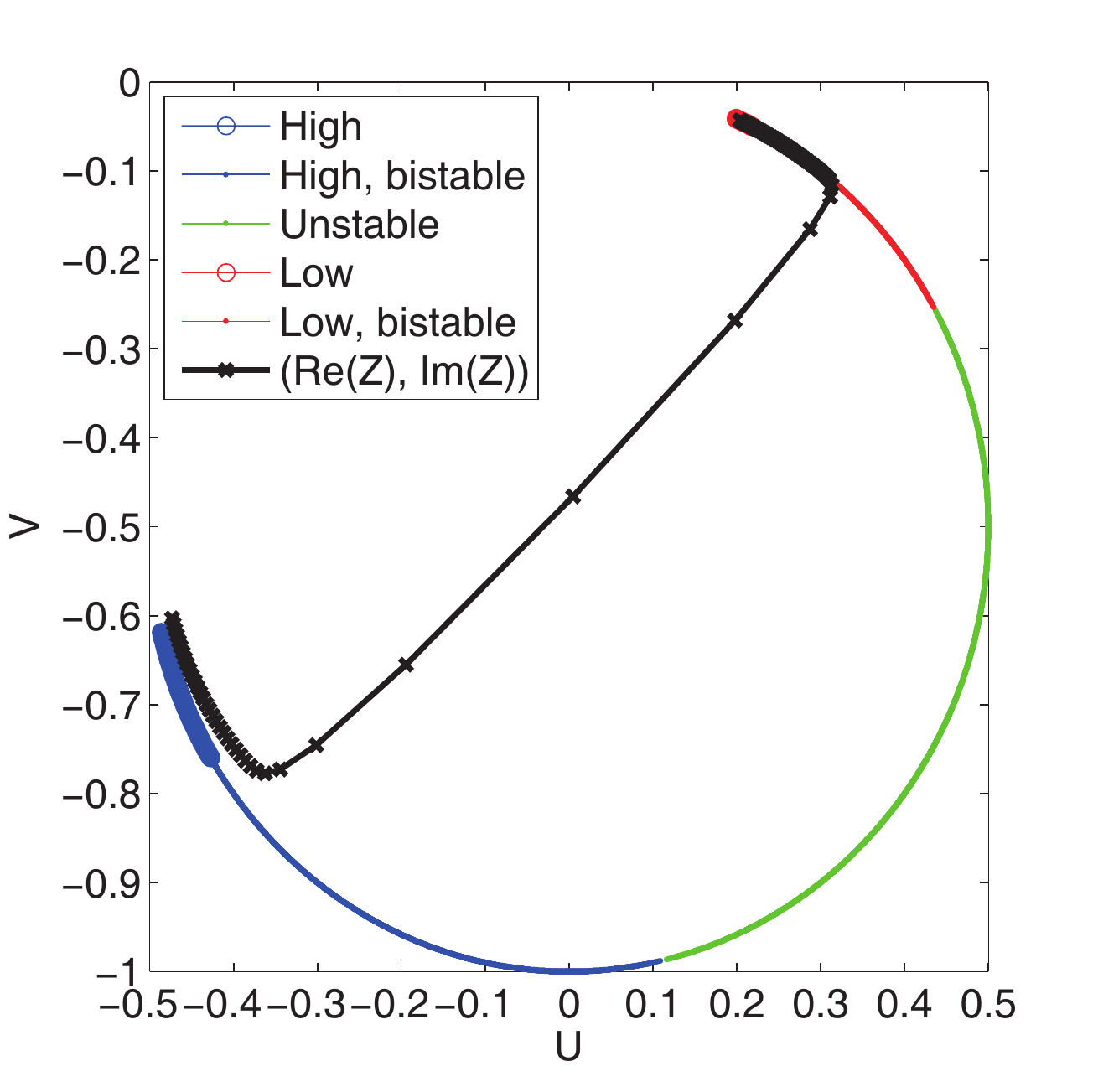}
  \caption{Amplitude crossover in the $(U,V)$ plane. Classical and quantum (for $f'=9$) data are parametric plots for $2\le\Omega'\le5$, the former using Eqs.~(\ref{classical-A}) and~(\ref{classical-phi}), and the latter using (\ref{def-Z}) for $Z'=U+iV$ [cf.\ above Eq.~(\ref{P-eq})]. Note that all classical stationary states fall on the circle $U^2+(V+\frac{1}{2})^2=\frac{1}{4}$, as is readily proven.}
  \label{complex-A}
\end{figure}

The correspondence to classical Duffing dynamics is vividly illustrated by the amplitude response (data points in Fig.~\ref{A-response}). To determine the proper expression for~$A'$, one can transform $\tilde{\rho}_1$ back to the lab frame using the inversion of Eq.~(\ref{RWA}) and calculate $\langle x\rangle(t)$ [the $t$-dependence arising solely from the one in~$U(t)$]. An unexpected feature is the small ``bump" visible on the low-amplitude side of the crossover; like the remainder of the crossover curves, this bump becomes narrower closer to the classical limit, but it retains a finite amplitude and never disappears. The issue is clarified by noting that in our driven system, the phase angle with respect to the driving force is relevant, and therefore one should properly introduce a \emph{complex} amplitude $Z\equiv Ae^{i\phi}$. Using the same considerations as above, one finds in dimensionless variables:
\be
  Z'=\frac{1}{f'}\sum_{n=0}^\infty\sqrt{2(n{+}1)}\langle n{+}1|\tilde{\rho}_1|n\rangle
  =\frac{\sqrt{2}}{f'}\Tr[\tilde{\rho}_1a]\;;\label{def-Z}
\ee
indeed, $A'$ in Fig.~\ref{A-response} was simply calculated as $\langle A'\rangle=|Z'|$. Thus, the amplitude crossover properly takes place in the complex plane, see Fig.~\ref{complex-A}. Amusingly, following the approximately straight crossover line formed by the data points from the low- to the high-amplitude branch initially brings one \emph{closer} to the origin, explaining the non-monotonous behaviour of $\langle A'\rangle(\Omega')$. Indeed, in a crossover plot of, say, the energy $\propto\langle N\rangle$, no such bump is observed. To be sure, this ``complex phase" phenomenon has nothing to do with quantum mechanics, and the amplitude response in the case of classical diffusion (\ref{FPE}) below features the ``bump" as well.

The approach to the classical response, where appropriate, provides one means of verifying our numerics. Direct comparison with analytical results is limited to the harmonic case~$\alpha'=0$, in which $\tilde{\mathcal{L}}$ can be diagonalized using superoperator Lie algebraic techniques~\cite{lie}. We omit the details, since this case is more easily dealt with in the coherent-state formulation (\ref{P-eq})--(\ref{Q-eq}) below. However, we do note that with further specialization to zero temperature, the ensuing stationary state $\tilde{\rho}_1$ is \emph{pure} even in the presence of dissipation, namely, a coherent state. The reason is that coherent states are eigenstates of the annihilation operator~$a$, which generates the only nonunitary process at $T=0$. All these are in full agreement with our numerical results.

\paragraph{Classical diffusion at finite temperature.}

The ultra-long tunneling time scale discussed above is only ``without classical counterpart" at temperature $T=0$. At $T>0$, noise-assisted transitions from one stationary state to the other are expected to occur, in analogy to thermal activation across the barrier of bistable potentials such as (\ref{dblwellmodel})~\cite{haenggi}. Thermal and quantum processes can be studied in combination by analyzing Eq.~(\ref{QME}) for finite $\bar{n}$; for now, we focus on thermal effects in the purely classical case~\cite{luck}.

Extending the Newton--Duffing dynamics into a diffusion process is most reliably done in the lab frame, by adding a Langevin term $\sqrt{2\epsilon\gamma k_\mathrm{B}T/m}\,\xi(t)$ to the rhs of Eq.~(\ref{ND}), with $\xi(t)$ being $\delta$-correlated Gaussian white noise. Equivalently, one can make the fluctuation--dissipation substitution $\epsilon\gamma\partial_pp\mapsto\epsilon\gamma\partial_p \bigl(p+mk_\mathrm{B}T\partial_p\bigr)$ in the associated Liouville (classical phase-space continuity) equation~\cite{opa}.

Transformation to the rotating frame can be done in either formulation, with equivalent results. In Langevin language, the $p$-noise in the lab frame becomes a rapidly rotating noise term in Van der Pol coordinates, which should be coarse-grained like the other evolution terms, keeping in mind that the noise autocorrelation time is the only time scale which is short compared to our ``fast" scale $\sim\omega^{-1}$. This results in separate, uncorrelated $u$- and $v$-noises to be added to Eq.~(\ref{coarse}). Alternatively, the diffusion term $\propto\partial_p^2$ can be transformed using Eq.~(\ref{uv}), upon which the coefficients of the resulting diffusion matrix are time-averaged separately. In either case, one obtains the Duffing--Fokker--Planck equation (FPE) with \emph{isotropic} diffusion term for the classical probability density~$\tilde{\rho}(U,V)$,
\begin{align}
  \frac{\partial\tilde{\rho}}{\partial\tau}&=-\bm{\nabla}\cdot(\bm{J}\tilde{\rho})
  +\frac{T_\mathrm{cl}}{2}\,\nabla^2\tilde{\rho}\label{FPE}\\
  &\equiv\tilde{\mathcal{L}}_{_\mathrm{FP}}\{\tilde{\rho}\}\;,\label{L-FP}
\end{align}
where $\bm{\nabla}=(\partial_U,\partial_V)^\mathrm{T}$ denotes the gradient with respect to the scaled coordinates of Eq.~(\ref{dim-less}), and with $\bm{J}$, $\tau$ as in Eq.~(\ref{coarse}). Instead of the reduced temperature $k_\mathrm{B}T/\hbar\omega_0$ entering Eq.~(\ref{QME}) through Eq.~(\ref{bose}), here one encounters the ``classical temperature" $T_\mathrm{cl}\equiv k_\mathrm{B}T\gamma^2/mf^2$, comparing $k_\mathrm{B}T$ to the energy $\sim m\omega_0^2A^2$ of classical oscillations with amplitude $A'=1$.

\begin{figure}
  \includegraphics[width=8.5cm]{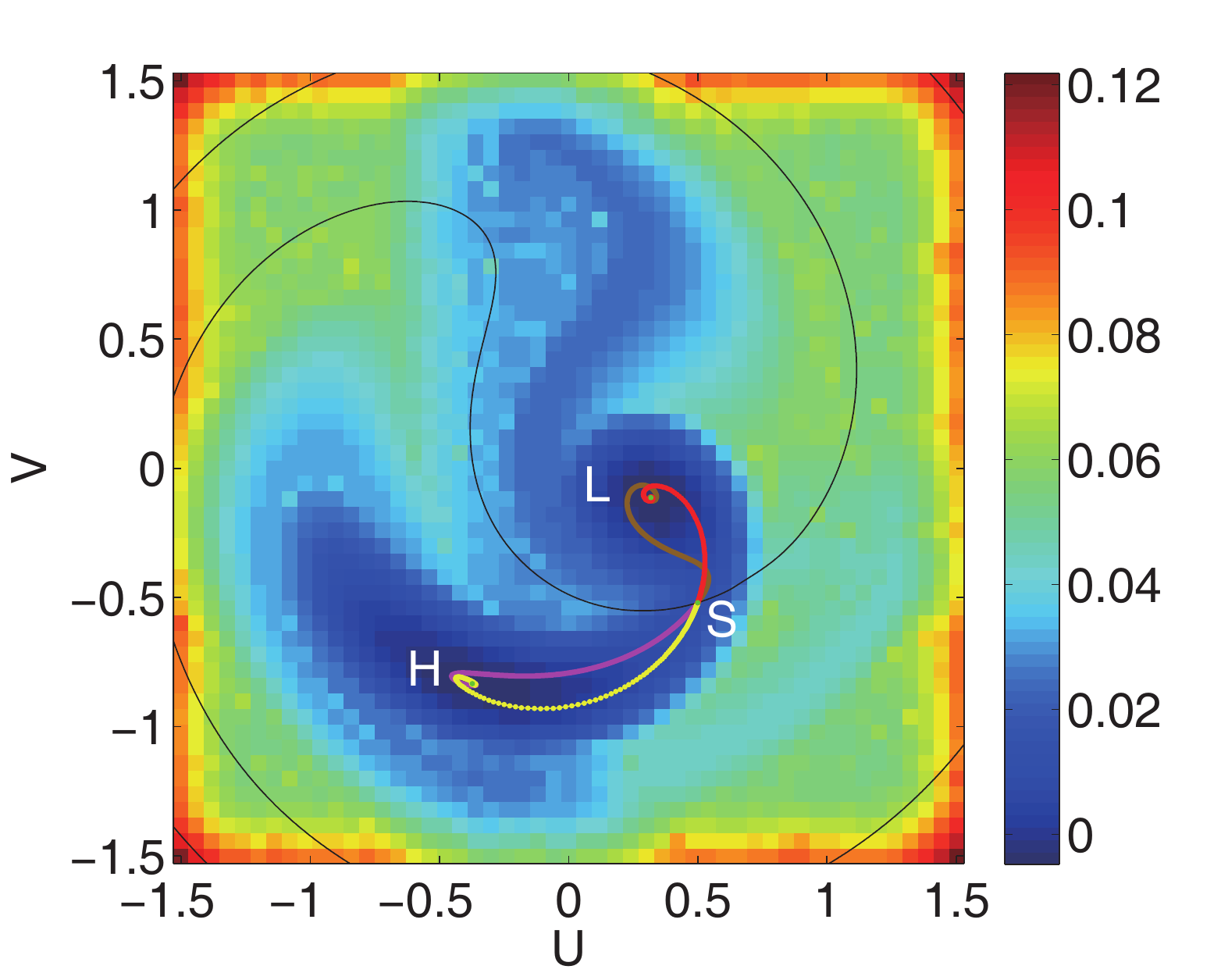}
  \caption{Shooting to the saddle point from the vicinity of the low- and high-amplitude stationary states, at detuning~$\Omega'=3.3$. Red/yellow: classical particle trajectories, integrated from the flow in Fig.~\ref{Jfig}. Black: separatrix. Magenta/brown: $\chi$-curves from Eq.~(\ref{shooting-eqn}), aimed towards the saddle to determine the effective barrier height. Background and colour scale: effective potential approximated via $-\ln\tilde{\rho}_1$ [cf.\ Eq.~(\ref{low-T})] for $T_\mathrm{cl}=0.0035$.}
  \label{FP}
\end{figure}

Equation (\ref{FPE}) can be studied by expanding it in a double Hermite basis~\cite{herm-scale},
\be
  \tilde{\rho}(U,V)=\sum_{nm}c_{nm}\psi_n^{(\ell)}(U)\psi_m^{(\ell)}(V)\;.\label{hermite}
\ee
Even though physically there are no creation and annihilation operators in classical diffusion, quantities such as $U$, $\partial_U$, etc are readily represented in matrix form using the recursion relations for Hermite functions, so that $\tilde{\mathcal{L}}_{_\mathrm{FP}}$ in Eq.~(\ref{L-FP}) is sparse similarly to $\tilde{\mathcal{L}}$ in~(\ref{L}). The resulting numerics are strongly analogous to those resulting from Eqs.~(\ref{QME})--(\ref{Hrot}) in the quantum case, and require resources of the same order~\cite{load}. As $T_\mathrm{cl}$ is lowered, the distribution $\tilde{\rho}(U,V)$ retreats to two narrow peaks centered on the stationary states (colour-scale component of Fig.~\ref{FP}), requiring increasingly many basis functions to represent accurately. Normalization of $\tilde{\rho}$ involves the doubly even coefficients $c_{2n,2m}$~\cite{FP-expand}. Extraction of the mean amplitude $\langle A'\rangle=\langle\sqrt{U^2+V^2}\rangle$ and of the probability distribution over the two peaks (for sufficiently low $T_\mathrm{cl}$ that such a quantity is well-defined) is comparatively less elegant, and can be performed on a grid by transforming back from matrix form to real space using Eq.~(\ref{hermite}).

Thus, the classical finite-temperature amplitude response and relaxation rate are obtained. As a function of decreasing~$T_\mathrm{cl}$, these are found to behave qualitatively similarly to the $T=0$ quantum data of Fig.~\ref{A-response} as a function of increasing $f'$. However, comparing the two sets of data in Fig.~\ref{extrapolate} reveals a curious difference: while each set individually is internally consistent (in that the ambiguity in the definition of the crossover point vanishes as the width of the crossover tends to zero), there is a small but significant difference in the value of the crossover point predicted by each set.

Since the only difference is that the QME data are at $T=0$ for $\hbar\To0$ while the FPE data are at $\hbar=0$ for $T\To0$, \emph{the system's behaviour in the $\hbar\To0$, $T\To0$ limit seems to depend on the order in which these two limits are taken}~\cite{thirdlimit}. If confirmed, this represents a striking difference from equilibrium tunneling in potentials such as~(\ref{dblwellmodel}), where crossover always occurs at potential-minimum degeneracy (even for asymmetric potentials, since zero-point motion in the well minima etc all vanish in the low-$T$ classical limit). The remainder of this paper is devoted to independent verification of this phenomenon, as well as to understanding of its origin.

\paragraph{Semi-analytical low-$T$ asymptotics.}

For an independent verification of what was obtained above through extrapolation of large-scale numerics, we have performed an asymptotic analysis. For the FPE (\ref{FPE}), this means attempting to take the low-$T$ limit analytically, through a quasi-Boltzmann Ansatz
\be
  \tilde{\rho}(U,V,T_\mathrm{cl})\sim e^{-2S(U,V)/T_\mathrm{cl}}\;,\label{low-T}
\ee
which plays a role here analogous to semi-classical optics or the WKB approximation in their respective fields. Substituting into the stationary limit of Eq.~(\ref{FPE}), to leading order in $T_\mathrm{cl}^{-1}$ one obtains
\be
  \bm{R}\cdot(\bm{R}+\bm{J})=0\;,\qquad\bm{R}\equiv\bm{\nabla}S\;.\label{RJ}
\ee
Thus, the original linear second-order PDE (\ref{FPE}) has been reduced to a nonlinear first-order one.

\begin{figure}
  \includegraphics[width=4cm]{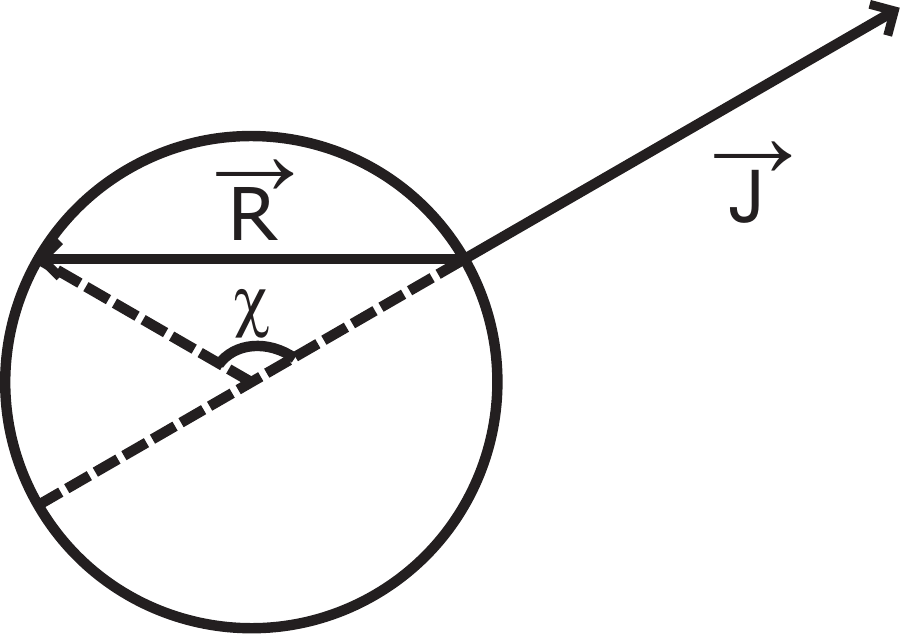}
  \caption{Equation~(\ref{RJ}) constrains $\bm{R}$ to a semicircle with diameter $-\bm{J}$.}
  \label{Rcircle}
\end{figure}

For a gradient flow $\bm{J}=-\bm{\nabla}U$, Eq.~(\ref{RJ}) is immediately solved by $S=U$, but for $\bm{J}$ as in Eq.~(\ref{Jflow}), the analysis is more involved. First, we solve Eq.~(\ref{RJ}) point-wise through the parametrization
\be
  \bm{R}=-\frac{\bm{J}}{2}+\frac{\bm{J}}{2}\cos\chi+\frac{\bm{J}_\perp}{2}\sin\chi\;,
\ee
with $\bm{J}_\perp\equiv(-J_V,J_U)^\mathrm{T}$ (see Fig.~\ref{Rcircle}). Instead of two unknown components of $\bm{R}$, one then deals with a single scalar field $\chi(U,V)$. To ensure that $\bm{R}$ can be uniquely integrated into an effective potential $S(U,V)$, one has to impose $\bm{\nabla}\times\bm{R}=0$ explicitly. Taking the curl of Eq.~(\ref{RJ}), this leads to
\be\begin{split}
  -[2\bm{R}(\bm{J},\chi)+\bm{J}]\cdot\bm{\nabla}\chi&=
  \Bigl(\frac{\partial J_U}{\partial V}-\frac{\partial J_V}{\partial U}\Bigr)
    (1-\cos\chi)\\
  &\quad+
  \Bigl(\frac{\partial J_U}{\partial U}+\frac{\partial J_V}{\partial V}\Bigr)\sin\chi\;.
  \label{shooting-eqn}
\end{split}\raisetag{5.5mm}\ee
Through the method of characteristic curves~\cite{smirnov}, this PDE is equivalent to a system of three coupled ordinary differential equations (for $U$, $V$, and~$\chi$), to which one may add the integration of $\bm{\nabla}S=\bm{R}(\bm{J},\chi)$ as a fourth for simultaneous solution.

The resulting differential system is highly nonlinear and unstable, with a parasitic solution $\chi=0\Rightarrow\bm{R}=0$. However, it has the key feature that the low-$T$ limit is already incorporated. To integrate a solution for $S$ one needs an initial value, which can be obtained at the stationary points of~$\bm{J}$, near which to leading order
\be
  \begin{pmatrix} J_U \\ J_V \end{pmatrix} \approx
  \begin{pmatrix} a & b \\ c & d \end{pmatrix}
  \begin{pmatrix} \delta U \\ \delta V \end{pmatrix}\;,
\ee
with different coefficients $a$ through $d$ for each of the three stationary points. Since one can convince oneself that these points are stationary also for $\bm{R}$ and for $\chi$, the rhs of (\ref{shooting-eqn}) can be set to zero there, yielding
\be
  \tan\frac{\chi_0}{2}=\frac{a+d}{c-b}\;.\label{chi0}
\ee
One cannot start \emph{at} a stationary point since there would be no evolution, but one can start close to it, taking $(\delta U,\delta V)=(\cos\theta,\sin\theta)\zeta$ with $\zeta\sim10^{-9}$--$10^{-6}$ (checking for convergence with respect to further decreasing~$\zeta$). The ultimate aim is to compare the values of $S$ in the low- and high-amplitude ``wells", which is achieved by a shooting method, tuning the angles $\theta$ such that one obtains curves linking either well to the saddle point, where for definiteness we set $S=0$ to fix the immaterial overall additive constant in~$S$. Implementing the method, we have found that different trajectories shot from the saddle to the wells converge exponentially, leading to loss of resolution. Thus, we start from the wells, and shoot out $\chi$-curves for a range of values for~$\theta$. Since the saddle is an unstable stationary point of these curves, $\theta$ must be tuned extremely accurately to approach the saddle closely; a representative case is shown in Fig.~\ref{FP}.

According to the above analysis, if the $\chi$-curves are indeed part of a global solution $S(U,V)$, then Eq.~(\ref{chi0}) should again be satisfied at the saddle (which has its own coefficients $a$ through~$d$). We have confirmed that this is indeed the case, for both low- and high-amplitude starting points; however, Eq.~(\ref{chi0}) is not satisfied at other points along the curve. Technically, the $\chi$-curves play a role similar to instanton paths in the WKB method; however, in the present setting we do not have a physical interpretation of these curves.

\begin{figure}
  \includegraphics[width=8.5cm]{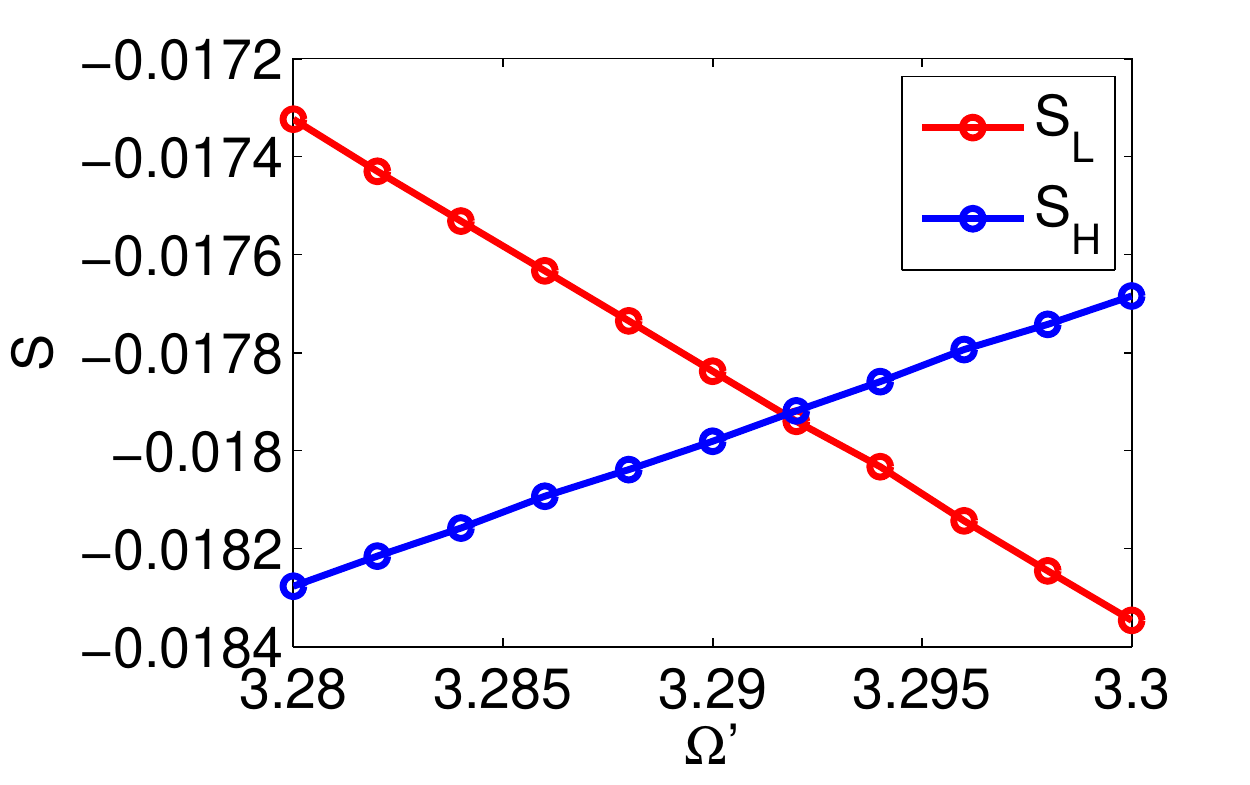}
  \caption{The effective potential $S$ versus detuning for the two classical stationary points. The two curves intersect at the crossover point.}
  \label{crossover-class}
\end{figure}

Carrying out the shooting procedure for a range of $\Omega'$ in the crossover region, we acquire the data shown in Fig.~\ref{crossover-class}. Reconstructing the effective potential $S(U,V)$ gives us our closest analogy with the potential bistability of Eq.~(\ref{dblwellmodel}). The inferred crossover point $\Omega'\approx3.292$ is shown as the red star in Fig.~\ref{extrapolate}. The excellent agreement with the extrapolated diagonalization data is compelling evidence that the latter indeed have the claimed accuracy (at least for the classical case). Note also that Fig.~\ref{crossover-class} predicts the high-amplitude state to dominate at lower detuning and vice versa, as is indeed observed in Fig.~\ref{A-response}.

Not only the crossover point, but also the corresponding barrier height $\Delta S$ is instructive. Namely, from an activation-energy consideration of Eq.~(\ref{low-T}), one expects $-\ln|\lambda_2|\sim2\Delta S/T_\mathrm{cl}+\text{const}$. Thus, $\Delta S$ can alternatively be inferred from the relaxation data $\lambda_2(T_\mathrm{cl})$ in a most traditional analysis---the Arrhenius plot~\cite{arrhenius}. We have verified that this yields $\Delta S\approx0.018\pm0.002$, in agreement with Fig.~\ref{crossover-class}. Thus, static ($\tilde{\rho}_1$) and dynamic ($\lambda_2$) data lead to consistent conclusions, as was also the case for the QME when inferring the incoherent nature of the tunneling either from $\re\lambda_2=0$, or from the absence of off-diagonal peaks in Fig.~\ref{rho-Nbasis}.

\paragraph{Coherent-state analysis.}

The above successful asymptotic study of the FPE (\ref{FPE}) prompts one to attempt a similar (large-$f'$) analysis for the QME (\ref{QME}) as well. However, the number (Fock) basis is notoriously unsuited for studying semiclassical properties of oscillator systems. Instead, it is logical to employ the ``most classical" oscillator states---the coherent states (in fact simply Gaussian wavepackets in the position basis) $|\zeta\rangle=e^{-|\zeta|^2/2}\sum_{n=0}^\infty\frac{\zeta^n}{\sqrt{n!}}|n\rangle$, with $\langle n\rangle=|\zeta|^2$. Due to their overcompleteness $|\langle\nu|\zeta\rangle|^2=e^{-|\nu-\zeta|^2}$, one has not one but two coherent-state representations:
\begin{align}
  \rho&=\int\!d^2\zeta\,P(\zeta)|\zeta\rangle\langle\zeta|\;,\label{def-P}\\
  Q(\zeta)&=\frac{1}{\pi}\langle\zeta|\rho|\zeta\rangle\;.\label{def-Q}
\end{align}

\begin{figure}
  \includegraphics[width=8.5cm]{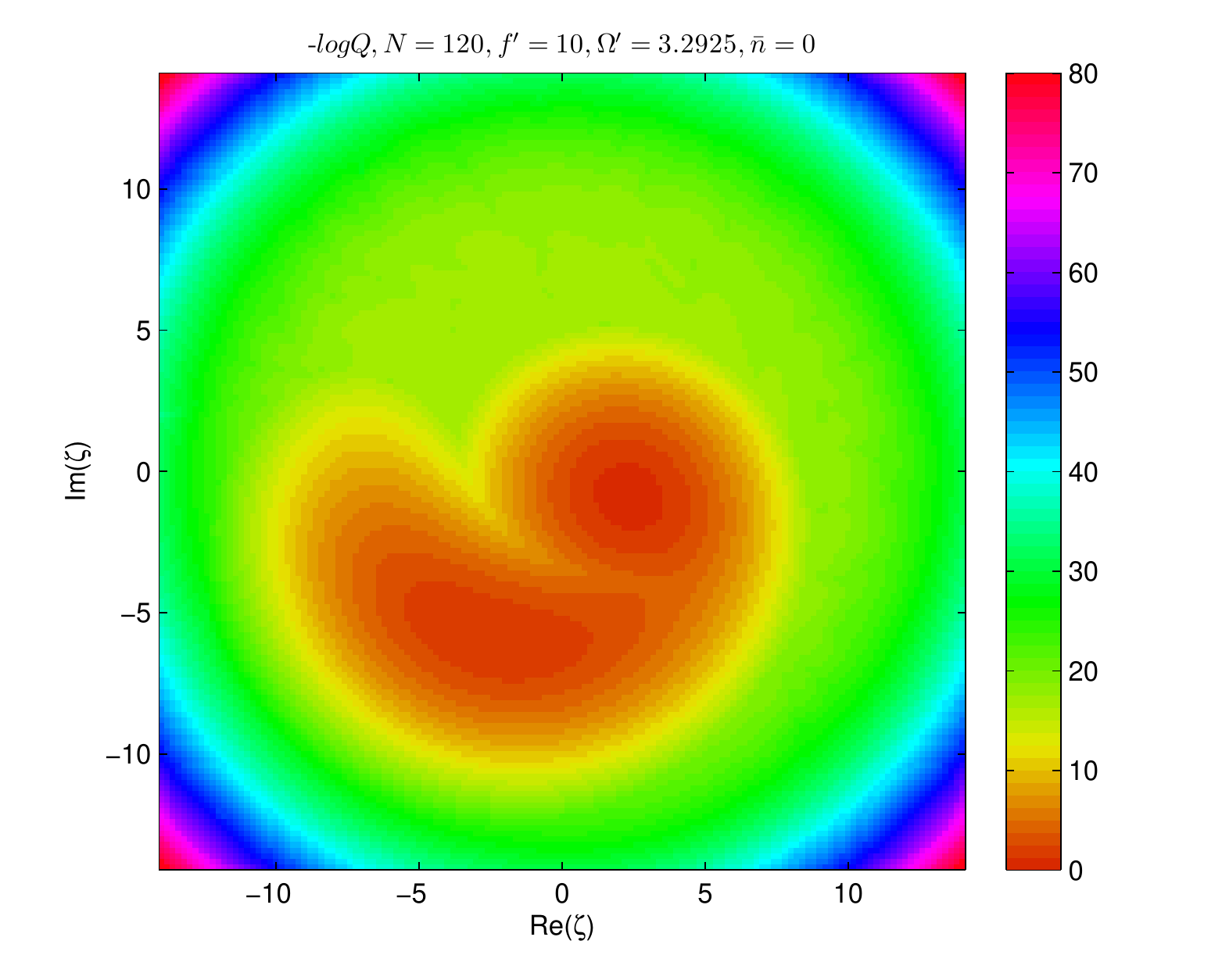}
  \caption{The coherent-state $Q$-function, obtained from its definition (\ref{def-Q}) for the stationary state $\tilde{\rho}_1$, close to the ($T=0$) crossover point.}
  \label{Q-function}
\end{figure}

Either one has its strengths and drawbacks. From the above definitions, one can immediately calculate $P\mapsto\rho\mapsto Q$. For our diagonalization data, $Q(U,V)$ is shown in Fig.~\ref{Q-function}; since it is \emph{real}, such a plot in principle characterizes the system's state fully, unlike Fig.~\ref{rho-Nbasis}, where the phase of the (off-diagonal) density-matrix elements is suppressed. Indeed, comparison of Figs.\ \ref{FP} and~\ref{Q-function} makes the classical--quantum correspondence manifest. The reverse operations $Q\mapsto\rho\mapsto P$ are more subtle. Some plausible-looking $Q$ are simply unphysical (i.e., not corresponding to any~$\rho$), for instance since Eq.~(\ref{def-Q}) endows $Q(\zeta)$ with a minimum linewidth. Also, while a $P$\nobreakdash-function can be proven to (uniquely) exist for any~$\rho$~\cite{harmonic-damping}, such functions are highly singular even for some simple $\rho$'s (e.g., pure number states).

Almost by definition, $a|\zeta\rangle=\zeta|\zeta\rangle$, and one also verifies $a^{\dagger}|\zeta\rangle=(\partial_\zeta+\zeta^*\!/2)|\zeta\rangle$. Thus, Eq.~(\ref{QME}) can be transcribed term by term into coherent-state language. In carthesian coordinates $(\re\zeta,\im\zeta)=(f'/\sqrt{2})(U,V)$, the results read
\begin{align}
  \frac{\partial P}{\partial\tau}&=-\bm{\nabla}\cdot(\bm{J}P)
    +\frac{\bar{n}}{2{f'}^2}\nabla^2P\notag\\
  &\quad+\frac{3\alpha'}{8{f'}^2}
    \bigl[(\partial_U^2-\partial_V^2)UV-\partial_U\partial_V(U^2-V^2)\notag\\
  &\hphantom{\quad+\frac{3\alpha'}{8{f'}^2}\bigl[}
    +2(\partial_VU-\partial_UV)\bigr]P\;,\label{P-eq}\raisetag{7mm}\displaybreak[0]\\
  \frac{\partial Q}{\partial\tau}&=-\bm{\nabla}\cdot(\bm{J}Q)
    +\frac{\bar{n}{+}1}{2{f'}^2}\nabla^2Q\notag\\
  &\quad-\frac{3\alpha'}{8{f'}^2}
    \bigl[(\partial_U^2-\partial_V^2)UV-\partial_U\partial_V(U^2-V^2)\notag\\
  &\hphantom{\quad+\frac{3\alpha'}{8{f'}^2}\bigl[}
    +2(\partial_VU-\partial_UV)\bigr]Q\;.\label{Q-eq}\raisetag{7mm}
\end{align}
The ``quantum terms" take a transparent form in polar coordinates,
\begin{multline}
  (\partial_U^2-\partial_V^2)UV-\partial_U\partial_V(U^2-V^2)+2(\partial_VU-\partial_UV)\\
  =-\partial_{A'}\partial_\phi A'\;.\label{q-terms}
\end{multline}
From the definitions (\ref{def-P}), (\ref{def-Q}) it follows that the $Q$\nobreakdash-function is merely a Gaussian smoothing of $P$, viz., $Q(\zeta)=\frac{1}{\pi}\int\!d^2\nu\,P(\nu)e^{-|\nu-\zeta|^2}$. Through integration by parts, this can be used to check the consistency of Eqs.\ (\ref{P-eq}) and~(\ref{Q-eq}).

In the harmonic case $\alpha'=0$ (note that $\alpha'$ also occurs implicitly in~$\bm{J}$), Eqs.\ (\ref{P-eq}) and~(\ref{Q-eq}) have a readily found Gaussian solution
\begin{align}
  P(U,V)&=\frac{1}{\pi\bar{n}}
    \exp\biggl\{-\frac{{f'}^2}{2\bar{n}}\mathcal{H}(U,V)\biggr\}\;,\\
  \mathcal{H}(U,V)&=\biggl(U-\frac{\Omega'}{1+{\Omega'}^2}\biggr)^{\!2}+
    \biggl(V+\frac{1}{1+{\Omega'}^2}\biggr)^{\!2}\;,
\end{align}
with $\bar{n}\mapsto\bar{n}+1$ for~$Q(U,V)$. In particular, for $\bar{n}\To0$ the $P$-function tends to a $\delta$-peak, which according to Eq.~(\ref{def-P}) corresponds to a coherent state [cf.\ below Eq.~(\ref{def-Z})].

The coherent-state equations (\ref{P-eq}), (\ref{Q-eq}) do not readily translate into a numerical scheme. While they would be satisfied by numerical solutions for $P$ and~$Q$ once found by other means, forward integration in $\tau$ is unstable since the ``quantum terms" (\ref{q-terms}) correspond to an indefinite ``diffusion matrix", as opposed to the positive definite~$\nabla^2$. More properly, the rhs of Eq.~(\ref{q-terms}) demonstrates that these terms represent a hyperbolic operator, as opposed to the elliptic~$\nabla^2$~\cite{smirnov}.

Instead, the strength of these equations is in making the various limits of interest more transparent. First of all, consider the classical limit at finite (classical) temperature. Since $\bar{n}/{f'}^2\approx T_\mathrm{cl}$ for $\bar{n}\gg1$, \emph{both} Eqs.\ (\ref{P-eq}) and~(\ref{Q-eq}) converge to Eq.~(\ref{FPE}) if $f'\To\infty$ at fixed finite $T_\mathrm{cl}$, the difference between them becoming negligible. Thus, classical diffusion is almost trivially obtained as a limit of the dissipative quantum dynamics. We have verified this behaviour numerically for both the stationary amplitude response and the tunneling rate (data not shown), using Eq.~(\ref{bose}) to determine the $\bar{n}(f',T_\mathrm{cl})$ to be used in Eq.~(\ref{QME})~\cite{finiteTstates}.

Contrast the above with the limit $f'\To\infty$ at fixed~$\bar{n}$ (implying $T_\mathrm{cl}\downarrow0$). In this case, both the (diffusive) $\nabla^2$\nobreakdash-term and the ``quantum terms" are $\mathcal{O}({f'}^{-2})$, so that neither dominates the other and both must be retained. The most striking case of this type of limit is $\bar{n}=0$; for the $P$-function, the diffusive term then is absent entirely, and only the ``quantum terms" cause a deviation from the classical contractive flow of Fig.~\ref{Jfig}. Thus, without actually having solved them, Eqs.\ (\ref{P-eq}) and~(\ref{Q-eq}) do yield the definite prediction that (and why) the limit ${f'}^{-1},T_\mathrm{cl}\To0$ depends on the value of $\bar{n}$, as was observed by various means above.

In an attempt to quantify these findings, for $\bar{n}=0$ we have pursued the asymptotic Ansatz analogous to Eq.~(\ref{low-T}):
\be
  P(U,V,f')\sim\exp\biggl[-\frac{8{f'}^2}{3\alpha'}\mathcal{H}(U,V)\biggr]\;.\label{P-asy}
\ee
Since it was pointed out above that the second-order differential operators in Eqs.\ (\ref{FPE}) and (\ref{P-eq}) have an essentially different character, and since we do not have independent numerical data for~$P$, such an approach is admittedly heuristic. Substitution of Eq.~(\ref{P-asy}) into (\ref{P-eq}) leads to a considerably more complicated analysis than the one leading up to Eq.~(\ref{shooting-eqn}) in the diffusive case---partly because of the position dependence in Eq.~(\ref{q-terms}), and partly because the counterpart to Eq.~(\ref{RJ}) constrains $\bm{\nabla}\mathcal{H}$ to hyperbolae not circles. We omit the details, since these considerations have been superseded by an unexpected analytic solution:
\be\begin{split}
  \mathcal{H}(U,V)&=-\frac{3\alpha'}{8}(U^2+V^2)+\frac{U}{U^2+V^2}\\
  &\quad+\frac{\Omega'}{2}\ln(U^2+V^2)-\phi\;,\label{Hexact}
\end{split}\ee
with $\phi$ the phase angle as above Eq.~(\ref{classical-A}).

\begin{figure}
  \includegraphics[width=8cm]{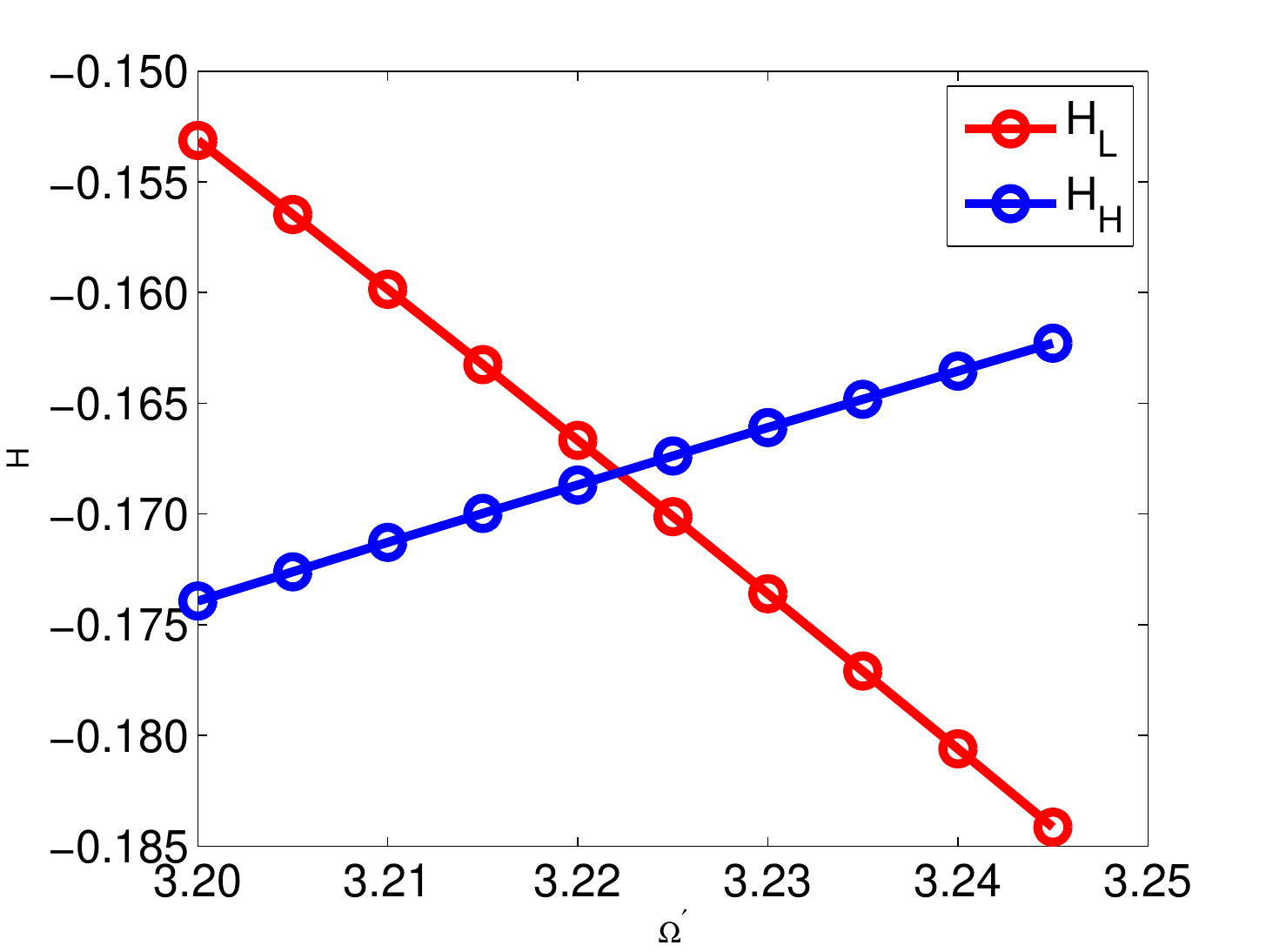}
  \caption{Crossover point for $\bar{n}=0$ (i.e., low-temperature limit first) as determined by degeneracy of the ``Quantum potential"~(\ref{Hexact}).}
  \label{qu-potential}
\end{figure}

One verifies that Eqs.\ (\ref{P-asy}), (\ref{Hexact}) solve the $\bar{n}=0$ coherent-state equation (\ref{P-eq}) not only to leading order in ${f'}^{-1}$, but in fact exactly. Construction of a crossover plot analogous to Fig.~\ref{crossover-class} merely involves the evaluation of the ``quantum potential" $\mathcal{H}$ at the classical stationary points, again using its value at the saddle point to fix the additive constant. The resulting data are shown in Fig.~\ref{qu-potential}.

It must be repeated that the whole procedure starting with Eq.~(\ref{P-asy}) is heuristic. We have not proven this asymptotic form for~$P$; given that the ensuing $\mathcal{H}$ has at least two major unphysical features (its multiple-valued character due to the occurrence of $\phi$ and its singularity at the origin), the Ansatz may indeed not be wholly correct. Nonetheless, the agreement of its upshot $\Omega'\approx3.222$ with the extrapolated numerical value for the $\bar{n}=0$ crossover point is unlikely to be mere coincidence. For the moment, we have to leave the solution's precise status as a subject of further study.

\paragraph{Conclusion.}

Let us review what has been accomplished. We have the two states visible in Fig.~\ref{Jfig}, whose definition requires nothing beyond 19th-century physics. The question which state is ``more stable" (or equivalently, for which parameter these states are ``equally stable") could be expected to be answerable in a similar setting. However, this intuition has been confounded; instead, the answer essentially depends on the details of how this limiting classical picture is reached from the underlying finite-temperature quantum theory. In contrast, for \emph{equilibrium} bistability described by Eq.~(\ref{dblwellmodel}), the corresponding determination can be made immediately from a plot, keeping in mind the word's literal meaning ``scales in balance".

It should be investigated whether other instances of nonequilibrium bistability~\cite{rolf} exhibit analogous phenomena and if so, whether a general description is possible. In any case, similar subtleties occur in several places in quantum physics; see, e.g., Ref.~\onlinecite{casimir} for a case of noncommuting zero-damping and low-temperature limits in the Casimir effect. Given that one needs extremely long times to reach the stationary state in the crossover regime [$\sim(\epsilon\lambda_2)^{-1}$, with $\lambda_2$ as in Fig.~\ref{separation}], one could verify whether going beyond the RWA affects any of our conclusions.

Moving beyond the specifics of the nonunique crossover, our investigations have yielded extensive data and calculation schemes for quantum Duffing oscillator, which is relevant to the qubit-readout application~\cite{JBA}. Acknowledging previous efforts~\cite{peano,saito,dykman,XQL}, the present results may be the most accurate and general [in the chosen regime of small $\epsilon$ in Eq.~(\ref{ND})].

Work supported by RCAS, Academia Sinica and the Taiwan National Science Council. AMB acknowledges the hospitality of the IPHT Jena, where some of the analytical work was done, and E.~Il'ichev (Jena), \mbox{W.-M.} Zhang (NCKU Tainan), H.-S. Goan (NTU Taipei), \mbox{X.-Q.} Li (Beijing Normal Univ.), L.~Tian (UC Merced) and Yu.V. Nazarov (Delft) for informative discussions, and M.~Nespoli for comments on the manuscript.


\begin{thebibliography}{99}

\bibitem{pedia} For a very accessible introduction, see\\ \texttt{www.scholarpedia.org/article/Duffing\_oscillator}

\bibitem{nori} See, e.g., J.Q. You and F.~Nori, Nature \textbf{474}, 589 (2011) for a recent review.

\bibitem{JBA} A. Lupa\c scu, S.~Saito, T.~Picot, P.C. de Groot, C.J.P.M. Harmans, and J.E. Mooij, Nat. Phys. \textbf{3}, 119 (2007).

\bibitem{saito} H. Nakano, S.~Saito, K. Semba, and H. Takayanagi, Phys. Rev. Lett. \textbf{102}, 257003 (2009).

\bibitem{dykman} M.I. Dykman, in \emph{Applications of Nonlinear Dynamics}, Understanding Complex Systems, edited by V.~In \emph{et al.}\ (Springer, Berlin, 2009), p. 367.

\bibitem{XQL} L. Guo, Z. Zheng, X.-Q. Li, and Y. Yan, Phys. Rev. E \textbf{84}, 011144 (2011).

\bibitem{peano} V. Peano and M.~Thorwart, Chem. Phys. \textbf{322}, 135 (2006).

\bibitem{alphanote} However, even if $\alpha$ were negative, tunneling out of the small-$|x|$ region (which is different from the tunneling phenomenon considered in this paper) would be exponentially weak in $\epsilon^{-1}$, and thus does not manifest itself in the regime considered. Instead, the $\alpha$-term mainly effects an amplitude-dependent shift in oscillation frequency.

\bibitem{VdP} B. van der Pol, Radio Review \textbf{1}, 701 (1920).

\bibitem{duffing-chaos} Y. Ueda, J. Stat. Phys. \textbf{20}, 181 (1979).

\bibitem{limitcycle} E.g., H.G. Schuster and W.~Just, \textit{Deterministic Chaos} (Wiley-VCH, Weinheim, 1995).

\bibitem{omega0} Since Eq.~(\ref{coarse}) already describes $\mathcal{O}(\epsilon)$ corrections to harmonic oscillation, the difference between $\omega$ and $\omega_0$ matters only to higher order in~$\epsilon$, so that the two variables can be used interchangably.

\bibitem{phi-sign} Equation (\ref{classical-phi}) determines $\phi$ only modulo~$\pi$. However, the correct solution always has $-\pi<\phi<0$ modulo~$2\pi$. Namely, the friction term $-\epsilon\gamma\dot{x}$ always lowers the system's energy, so that energy balance in the stationary state requires the driving force to perform work on the system on average, fixing the sign of the phase angle $\phi$ between this force and the oscillations in~$x(t)$.

\bibitem{driven-tunnel} U.~Weiss, \textit{Quantum Dissipative Systems} (World Scientific, Singapore, 1993).

\bibitem{zhang} J.~Jin, M.~Tu, W.-M. Zhang, and Y.~Yan, New J. Phys. \textbf{12}, 083013 (2010) and references therein.

\bibitem{harmonic-damping} C.W. Gardiner and P.~Zoller, \textit{Quantum Noise} (Springer, Berlin, 2004).

\bibitem{lindblad} V. Gorini, A. Kossakowski, and E.C.G. Sudarshan, J.~Math. Phys. \textbf{17}, 821 (1976); G. Lindblad, Commun. Math. Phys. \textbf{48}, 119 (1976).

\bibitem{JB} In the non-hermitian case, there is no guarantee that a given matrix can be diagonalized at all [e.g., V.A. Ilyin and E.G. Poznyak, \textit{Linear Algebra} (Mir, Moscow, 1986)]. However, exceptions are expected to be limited to isolated points in parameter space only.

\bibitem{phase-rel} The two lobes of $\tilde{\rho}_1$ each have a definite phase (which can be characterized in detail using standard quantum-optical methods, but which is already apparent from Fig.~\ref{complex-A}), and thus there is a definite phase relationship between them. However, this is because both lobes are synchronized to the same external clock---the driving force---and \emph{not} due to the tunneling transitions between them.

\bibitem{lie} B.P. Hou and S.J. Wang, Phys. Lett. A \textbf{311}, 106 (2003).

\bibitem{haenggi} P. H\"anggi, P. Talkner, and M. Borkovec, Rev. Mod. Phys. \textbf{62}, 251 (1990).

\bibitem{luck} Initially, the Fokker--Planck formulation was studied in an attempt to find a simplified expression for the---supposedly unique---crossover point. Our main nonuniqueness finding was serendipitous.

\bibitem{opa} N.G. van Kampen, \textit{Stochastic Processes in Physics and Chemistry} (North-Holland, Amsterdam, 1981).

\bibitem{herm-scale} Unlike in the quantum case, one here has the freedom to choose the length scale $\ell$ of the basis functions $\psi_n^{(\ell)}(U)\equiv\ell^{-1/2}\psi_n(U/\ell)$, with $\psi_n(x)=(n!2^n\sqrt{\pi})^{-1/2}e^{-x^2/2}H_n(x)$ being the standard orthonormal Hermite functions; $H_n(x)=(-)^ne^{x^2}d_x^ne^{-x^2}$ is the $n$th Hermite polynomial. Judicious choice of $\ell$ is vital for rapid convergence. We find that $\ell$ should be decreased as more basis functions are being included; the reason is that one needs higher-order Hermite functions not to describe features at larger $U$ and $V$, but to resolve increasingly fine details on the fixed scale $|U|,|V|\sim1$.

\bibitem{load} However, the computational load in the quantum case grows without bound as the classical limit is approached. Also, note that the matrix for $\tilde{\mathcal{L}}_{_\mathrm{FP}}$ is purely real, which enables one to use a more economical diagonalization routine than for~$\tilde{\mathcal{L}}$.

\bibitem{FP-expand} Sometimes an alternate expansion $\tilde{\rho}(U,V)=\sum_{nm}d_{nm}\psi_0^{(\ell)}(U)\psi_n^{(\ell)}(U)
    \psi_0^{(\ell)}(V)\psi_m^{(\ell)}(V)$ is used, with the advantage that the normalization of $\tilde{\rho}$ involves $d_{00}$ only due to the orthogonality of the~$\psi_n$ [H.~Risken, \textit{The Fokker--Planck Equation} (Springer-Verlag, New York, 1984)]. However, such an expansion complicates the form of~$\tilde{\mathcal{L}}_{_\mathrm{FP}}$; since the analysis of the latter is the limiting step, we have instead proceeded using Eq.~(\ref{hermite}) in the main text.

\bibitem{thirdlimit} This subtlety only occurs---and can only occur---because we consider the classical, low-temperature limit of the stationary (or at least, ultra-long-time asymptotic relaxation) data, i.e., after the $\tau\To\infty$ limit has already been taken. If $\hbar\To0$ and $T\To0$ are taken (in any order) at \emph{finite} time, one simply recovers the hysteretic classical dynamics of Eq.~(\ref{coarse}).

\bibitem{smirnov} W.I. Smirnow, \textit{Lehrgang der h\"oheren Mathematik}, Teil IV/2 (VEB Deutscher Verlag der Wissenschaften, Berlin, 1988).

\bibitem{arrhenius} S. Arrhenius, Z. Phys. Chem. (Leipzig) \textbf{4}, 226 (1889).

\bibitem{finiteTstates} Since both the driving force and thermal noise can excite the oscillator, numerics at $T>0$ require even more photon states to ensure convergence.

\bibitem{rolf} R. Landauer, J. Stat. Phys. \textbf{53}, 233 (1988).

\bibitem{casimir} G.-L. Ingold, A.~Lambrecht, and S.~Reynaud, Phys. Rev. E \textbf{80}, 041113 (2009).

\end{thebibliography}
\end{document}